\documentclass[aps,prx,twocolumn,showpacs,superscriptaddress]{revtex4-2}
\usepackage{graphicx}  
\usepackage{dcolumn}   
\usepackage{bm}        
\usepackage{dsfont}    
\usepackage{amssymb}   
\usepackage{amsmath}
\usepackage{color}
\usepackage{siunitx}
\usepackage{physics}
\usepackage{hyperref}
\usepackage{xcolor}

\usepackage[utf8]{inputenc}

\definecolor{qred}{HTML}{ac2b37} 
\definecolor{hawaii}{HTML}{00452a} 
\newcommand{\chg}[1]{{#1}}

\newcommand{\powerset}{\raisebox{.15\baselineskip}{\Large\ensuremath{\wp}}}

\DeclareRobustCommand{\okina}{%
	\raisebox{\dimexpr\fontcharht\font`A-\height}{%
		\scalebox{0.8}{`}%
	}%
}
\newcommand{\hawaii}{Hawai\okina i}

\begin{document}
	\title{Statistical Mechanics of Dynamical System Identification}
	\author{Andrei A.~Klishin}

        \email{aklishin@uw.edu}
        \altaffiliation[Current address: ]{Department of Mechanical Engineering, University of \hawaii~at M\=anoa, Honolulu, HI 96814, USA}
	\affiliation{AI Institute in Dynamic Systems, University of Washington, Seattle, WA 98195, USA}
	\affiliation{Department of Mechanical Engineering, University of Washington, Seattle, WA 98195, USA}

	\author{Joseph Bakarji}

	\affiliation{AI Institute in Dynamic Systems, University of Washington, Seattle, WA 98195, USA}
	\affiliation{Department of Mechanical Engineering, University of Washington, Seattle, WA 98195, USA}
	\affiliation{Department of Mechanical Engineering and Artificial Intelligence Hub, American University of Beirut, Beirut, 1107 2020, Lebanon}

	\author{J.~Nathan Kutz}
	\affiliation{AI Institute in Dynamic Systems, University of Washington, Seattle, WA 98195, USA}
	\affiliation{Departments of Applied Mathematics and Electrical and Computer Engineering, University of Washington, Seattle, WA 98195, USA}

	\author{Krithika Manohar}
		\email{kmanohar@uw.edu}
	\affiliation{AI Institute in Dynamic Systems, University of Washington, Seattle, WA 98195, USA}
	\affiliation{Department of Mechanical Engineering, University of Washington, Seattle, WA 98195, USA}

	\date{\today}
	
	\begin{abstract}
		Recovering dynamical equations from observed noisy data is the central challenge of system identification. We develop a statistical \chg{mechanics} approach to analyze sparse equation discovery algorithms, which typically balance data fit and parsimony via hyperparameter tuning. In this framework, statistical mechanics offers tools to analyze the interplay between complexity and fitness \chg{similarly to that of} entropy and energy \chg{in physical systems}. To establish this analogy, we define the hyperparameter optimization procedure as a two-level Bayesian inference problem that separates variable selection from coefficient \chg{inference} and enables the computation of the posterior parameter distribution in closed form. \chg{Our approach provides uncertainty quantification, crucial} in the low-data limit that is frequently encountered in real-world applications. A key advantage of employing statistical mechanical concepts, such as free energy and the partition function, is \chg{to connect the large data limit to thermodynamic limit and characterize the} sparsity- and noise-induced phase transitions that delineate correct from incorrect identification. \chg{We thus provide a method for closed-loop inference, estimating the noise in a given model and checking if the model is tolerant to that noise amount.} This perspective of sparse equation discovery is versatile and can be adapted to various other equation discovery algorithms.
        
	\end{abstract}
	
	\maketitle

\section{Introduction}
Identifying dynamical models from data represents a critical challenge in a world inundated with emerging time-series data which lack accurate and robust characterizations. This is the central task of modern system identification \cite{ljung2010perspectives}. Traditional methods of constructing ordinary and partial differential equations (ODEs and PDEs) from the first principles are increasingly limited by our intuitive understanding and the growing complexity of contemporary problems involving high-dimensional data and unfamiliar nonlinearities \cite{north2023review}. Conversely, modern deep learning approaches, while capable of identifying highly nonlinear relationships from time-dependent data \cite{li2020fourier, rudin2022interpretable}, often struggle with overfitting and lack the interpretability essential for human element in improving model generalization \cite{lipton2018mythos, rudin2019stop}.

This backdrop sets the stage for modern data-driven methods that search through a large set of hypothetical differential equations to achieve an optimal fit with observational data \cite{crutchfield1987equations}. Notable examples include Sparse Identification of Nonlinear Dynamics (SINDy)~\cite{brunton2016discovering} and its extensions~\cite{rudy2017pde, kaiser2018sparse, messenger2021weak, hokanson2023simultaneous}, Symbolic Regression~\cite{schmidt2009distilling, udrescu2020aifeynman}, Sir Isaac~\cite{daniels2015automated}, and equation learning~\cite{sahoo2018learning}. These methods have seen diverse applications from sparse biochemical reaction networks \cite{mangan2016networks} to atmospheric chemistry surrogate modeling, \cite{yang2024atmospheric}, uncertainty quantification~\cite{bakarji2021data}, active matter~\cite{supekar2023learning}, and fluid dynamics~\cite{loiseau2018constrained}. The parsimonious form of equations, apart from being directly interpretable by domain experts, also enables the key physical features of the model, namely generalization and extrapolation \cite{kutz2022parsimony}.

Despite their promise, these techniques face the added complexity of balancing parsimony with accuracy through trial-and-error hyperparameter tuning \cite{vanbreugel2020numerical}. The effectiveness of these methods is contingent on the amount of available data and the level of noise present. There is currently a lack of understanding regarding these limiting cases, and how the various hyperparameters interact with them. \chg{The inferred model is naturally a function of the chosen hyperparameters, usually the prior parameters or the coefficients of loss function terms such as sparsity penalty. The two most common strategies are to either fix the hyperparameters at some values that happen to work well for the example dataset \cite{gao2022bayesian, north2022bayesian, fung2024rapid}, or "optimize away" the hyperparameters through cross-validation \cite{vanbreugel2020numerical, wentz2023derivative}. Both of these strategies obscure the inherent trade-off between accuracy and parsimony, which might be useful for an applied practitioner but is contrary to the scientific method.}

In this paper, we \chg{directly expose the accuracy---parsimony trade-off} using \chg{\emph{exact}} Bayesian inference within a statistical mechanical approach to sparse equation discovery. Statistical mechanics has long been used to analyze the average-case behavior of inference problems from classifiers to sparse sensing and network structure \cite{langley1999tractable, ganguli2010statistical, nadakuditi2012graph, zdeborova2016statistical, obuchi2018statistical, krzakala2022notes}. \chg{Those studies primarily focus on a large number of features that have different collective inference behaviors but are individually unimportant and anonymous; in contrast,} here we apply the same ideas to identify a small set of mechanistically interpretable, non-anonymous terms of dynamical equations. 

The proposed method, Z-SINDy, \chg{incorporates two key advancements: \emph{fast closed-form computations of the full posterior distribution} and \emph{direct $L_0$ sparsity penalty} as opposed to its approximations via $L_1$, iterative reweighing, or sequential thresholding.} The key computational advantage lies in separating the accounting of the discrete and continuous degrees of freedom similar to mixed-integer optimization \cite{bertsimas2023learning}, and using a closed-form expression for multivariate Gaussian integrals. \chg{Many SINDy-inspired studies in the recent years have taken an \emph{approximate} Bayesian view to provide error bars of the inferred models through methods such as Monte Carlo or Gibbs sampling \cite{hirsh2022sparsifying, north2022bayesian, gao2022bayesian, 1wang2020perspective, 3mathpati2024discovering, 4tripura2023sparse, 6nayek2021spike}, Variational Bayes \cite{3mathpati2024discovering, 5more2023bayesian, 7long2024equation, 8niven2020bayesian}, or Expectation-Propagation Expectation-Maximization \cite{7long2024equation}. The methods most similar to ours are presented in Refs.~\cite{fung2024rapid, niven2024dynamical}, but those studies do not analyze how the confidence intervals of the inferred model parameters scale with data size or noise magnitude, or how the sparsity promotion generates a family of models that trade off complexity and fitness.}

\chg{While some system identification studies provide guarantees of convergence to the correct model in the low noise limit \cite{10messenger2022asymptotic}, the numerical validation experiments typically do not report failure cases where inference fails in the sense of finding qualitatively wrong equation terms \cite{zhang2019convergence, hokanson2023simultaneous, 12tran2017exact, 13schaeffer2018extracting, 14schaeffer2020extracting}}. At the same time, even when uncertainty quantification is provided for a few particular datasets and noise levels, there is no general argument on how both the width of error bars and the systematic error in the coefficients scale with dataset parameters \cite{zhang2018robust}. In other words, when identification fails, we do not know why. \chg{In this paper we aim to primarily figure out under what conditions the identified solution would match the ground truth. While computational efficiency was not the primary goal, in many practical cases Z-SINDy computation is much faster than alternative methods.}


When Z-SINDy is applied to low-data scenarios, it returns the full posterior distribution over the dynamical models and thus provides uncertainty quantification at a fraction of computational cost of other approaches \cite{fasel2022ensemble, hirsh2022sparsifying, north2022bayesian, gao2022bayesian, gao2023convergence}. In high-data scenarios, we show that inference always condenses to a definite, though not necessarily correct model, and switches abruptly between models. Accordingly, as either the noise in the dataset or the sampling period are increased, we observe a detectability phase transition from correct to incorrect dynamical model, directly informing the trade-off between model fidelity and sparsity. \chg{The uncertainty quantification and noise tolerance metrics enable closed-loop inference, that is finding a model and checking if that model could be found given the amount of noise.}
The statistical mechanics analysis can be further integrated with other SINDy advancements or applied to other cases of sparse inference.

\section{Statistical Mechanics for Sparse Inference}
\begin{figure*}
	\centering
	\includegraphics[width=.8\textwidth]{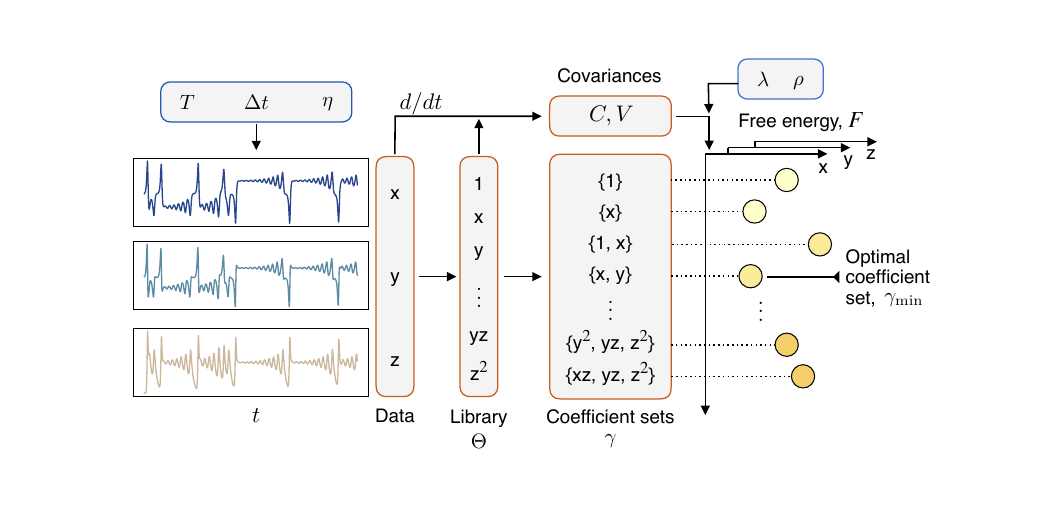}
	\caption{Schematic representation of the questions Z-SINDy aims to answer. A dataset consists of a trajectory of time length $T$ sampled with time step $\Delta t$ and measurement noise of magnitude $\eta$ (left blue box). From the data, a library of nonlinear functions $\Theta_i(\vec{x})$ is \chg{computed}. The library functions are combined with empirical derivatives of the trajectory to compute the covariance matrices $C$ and vectors $\vec{V}$. The covariances along with the algorithm hyperparameters of sparsity $\lambda$ and resolution $\rho$ (right blue box) are used to compute the free energies \chg{$F^{(l)}_{\gamma_l}$} that quantify the balance between goodness of fit and number of fitting solutions. The best fit coefficient set $\gamma_{min}$ is the one that corresponds to the lowest free energy. In this paper we study how the probability of choosing the correct coefficient set depends on the parameters in blue boxes.}
	\label{fig:setup}
\end{figure*}

\subsection{Background}
System identification begins with an observed trajectory $\vec{x}(t)$ of a $d$-dimensional dynamical system. The trajectory covers the length of time $T$ sampled with period $\Delta t$, resulting in $n=T/\Delta t$ data points. 
For synthetic trajectory data, we presume that the integration time step is much smaller than $\Delta t$, making the integration error negligible. 
Trajectory measurement incurs uncorrelated additive Gaussian noise of magnitude $\eta$ in each dimension of dynamics. From the trajectory we compute the empirical derivative $\dot{\vec{x}}$ with a second-order centered finite difference method to simplify the interpretation.

The goal of sparse equation discovery is to extract a dynamical equation $\dot{\Vec x} = f(\Vec x)$ from the observed trajectories of $\Vec x$, where $f(\cdot)$ is a sparse analytical expression. When $f(\cdot)$ is given by a library of candidate nonlinear functions, we seek an equation of the form:
\begin{align}
	\dot{x_l}\approx \sum\limits_{i=1}^{N} \Theta_i(\Vec{x})^T\Xi_{il},
	\label{eqn:eom}
\end{align}
where the index $l$ enumerates the dynamical variables, the left hand side is the empirical derivative and right hand side is a linear combination of $N$ nonlinear functions of dynamical variables $\Theta_i(\Vec{x})$ (i.e. the library) with a matrix of coefficients $\Xi_{il}$. The Sparse Identification of Nonlinear Dynamics (SINDy) seeks to optimize for the \emph{sparsest} matrix of coefficients $\Xi_{il}$ that minimizes the residual of~\eqref{eqn:eom}. This optimization problem involving two competing objectives - sparsity and fitness - leads to a challenging non-uniqueness in the resulting models.
In practice, the problem is typically solved by optimizing a loss function that combines linear regression with a sparsity penalty. While sparsity is directly measured by the $\norm{\Xi}_0$ pseudonorm on the model coefficients, it is usually approximated by either the $\norm{\Xi}_1$ norm or sequential thresholded least squares \cite{donoho2006most, brunton2016discovering}. 

Where optimization-based SINDy provides a point estimate of the coefficients, Bayesian approaches aim to extract the maximum amount of information from the data, while providing an uncertainty quantification of the resulting equation. In the Bayesian setting, sparsity can be promoted by the choice of a sparsifying prior, such as Laplace, spike-and-slab, or horseshoe \cite{carvalho2009handling, hirsh2022sparsifying, gao2022bayesian}. These priors aim to concentrate the posterior probability near $\Xi_{il}=0$, while remaining differentiable to enable Monte Carlo sampling.

The interplay of sparsity with large numbers of variables and data points in a probabilistic setting attracted significant attention from the statistical mechanics community, particularly the theory of disordered systems \cite{krzakala2022notes}. By using the so-called replica trick, researchers averaged over random data matrices to obtain the average behavior of different classes of inference problems, revealing multiple detectability and algorithmic phase transitions \cite{ganguli2010statistical, zdeborova2016statistical, obuchi2018statistical, bereyhi2020statistical}. Our approach here is different in two key aspects: first, we focus on identifying not a finite \emph{fraction} of relevant variables but a finite \emph{number}, thus attaching more interpretation to each term; second, instead of averaging over generic random Gaussian data matrices, we work with trajectory data, including the sampling period and numerical differentiation effects.

\subsection{Z-SINDy}
In the present paper we separate the Bayesian inference problem \chg{of coefficients $\Xi$} into two layers: the discrete layer describes which \chg{set of} coefficients \chg{$\gamma$ is} active (non-zero), \chg{e.g. $\gamma=\varnothing,\{\text{x,yz}\},\{\text{1,y,z,xy,xz}\}$}, while the continuous layer describes the values of coefficients \chg{$\Xi_\gamma$ in the active set}. \chg{The inference framework is schematically represented in Fig.~\ref{fig:setup}}.

In particular, this new setting does not require the prior to be differentiable, allowing us to use a much  simpler Bernoulli-Gaussian functional form \cite{obuchi2018statistical}:
\begin{align}
	p(\Xi)=& \prod_i \frac{1}{1+e^{-\Lambda}}(\delta(\Xi_{i})+w(\Xi_{i})e^{-\Lambda})\label{eqn:prior}\\
	=&\frac{1}{(1+e^{-\Lambda})^N} \sum\limits_\gamma \left( \prod\limits_{i\notin \gamma} \delta(\Xi_i) \right) \left( \prod\limits_{i\in \gamma} w(\Xi_i) \right) e^{-\Lambda \abs{\gamma}},
	\label{eqn:priorfact}
\end{align}
where we omitted the index $l$ for simplicity of notation, since the inference is independent in each dimension with a shared set of library terms $\Theta_i(\vec{x})$. Within the Bernoulli-Gaussian prior, the hyperparameter $\Lambda$ regulates sparsity; $\delta(\cdot)$ is Dirac delta function that can be thought of as an infinitely narrow Gaussian; and $w(\cdot)$ is a prior function that quantifies the parameter uncertainty in the absence of data, which we take to be in the limit of an infinitely wide Gaussian (see SI for additional discussion \chg{of the prior}).

The prior form of \eqref{eqn:prior} is symmetric under the permutation of indices $i$, highlighting that the inference procedure does not have a preference of any variable combination before seeing any data. The symmetry is made even more explicit by multiplying out the brackets to get to \eqref{eqn:priorfact}, which is given as a \chg{combinatorial} sum over the \emph{coefficient sets} \chg{$\gamma$}. For $N$ library terms there are $2^N$ possible discrete coefficient sets that form the power set of library indices $\{\gamma_l\}=\powerset(\{i\})$. The prior probability of any coefficient set depends only on its size $\abs{\gamma}$ but not its identity, and increasing the value of $\Lambda$ shifts progressively more probability weight from larger to smaller sets, while maintaining unbroken symmetry under index permutation.

Along with the prior, we define the forward model, i.e. the probability of observing the data given the coefficients (likelihood):
\begin{align}
	p(x|\Xi)\propto \exp(-\frac{1}{2\rho^2}\sum\limits_t \left( \dot{x}-\vec{\Theta}^T \vec{\Xi} \right)^2 ),
	\label{eqn:forward}
\end{align}
where we suppressed the time indexing on the empirical derivative and the library terms. The resolution hyperparameter $\rho$ regulates how closely the nonlinear function library aims to approximate the empirical dynamics derivative.

Using the forward model and the prior, we derive the posterior via Bayes rule:
\begin{align}
	p(\Xi|x)=\frac{1}{\mathcal{Z}} \exp(-\frac{1}{2\rho^2}\sum\limits_t \left( \dot{x}-\vec{\Theta}^T\vec{\Xi} \right)^2 -\Lambda \norm{\vec{\Xi}}_0),
	\label{eqn:posterior}
\end{align}
where the expression inside $\exp(\cdot)$ is the \chg{$L_0$} regularized SINDy functional \cite{brunton2016discovering}. In statistical mechanics, this would be akin to Boltzmann's distribution with $\rho^2$ playing the role of temperature, $\Lambda$ playing the role of chemical potential (cost of adding new particles to the system), and $\mathcal{Z}$ the partition function (normalization) of the distribution. The mode of this distribution, or the Maximum A Posteriori (MAP) estimate of the coefficients, would be equivalent to the original SINDy problem statement, but the \chg{$L_0$} regularization is challenging to optimize in practice.

Instead of performing optimization, we disentangle the mixture of continuous and discrete degrees of freedom in \eqref{eqn:posterior} by using the factorized form of the prior \eqref{eqn:priorfact} to rewrite the posterior into a hierarchical form as a choice of active coefficient set followed by a choice of coefficient values. In this factorized form the posterior takes form of a linear mixture of multivariate Gaussians:
\begin{align}
	p(\Xi|x)=&\frac{\sum_\gamma p(\vec{\Xi}_\gamma|\gamma,x) \mathcal{Z}_{\gamma}}{\sum_{\gamma} \mathcal{Z}_{\gamma}}
	\label{eqn:posteriorfact}\\
	\mathcal{Z}_{\gamma}\equiv& e^{-F_{\gamma}} = e^{-F'_{\gamma}} e^{-\Lambda\abs{\gamma}}\\
	F'_{\gamma}=& -\ln \mathcal{Z}_0 -\frac{|\gamma|}{2}\ln(2\pi \rho^2)+ \frac{1}{2} \ln \det C_{\gamma} \nonumber \\
	-& \frac{1}{2\rho^2} \vec{V}^T_{l,\gamma} C^{-1}_{\gamma} \vec{V}_{l,\gamma_l}
	\label{eqn:Fgamma}\\
	F_{\gamma}\equiv& F'_{\gamma}+\Lambda \abs{\gamma}=F'_{\gamma}+\lambda\cdot n \abs{\gamma},
	\label{eqn:Fgammaprime}
\end{align}
where the \emph{statistical weights} $\mathcal{Z}_{\gamma}$ quantify the relative importance of each coefficient set (evidence) and are thus the central objects of the method, inspiring the name Z-SINDy. Since the statistical weights vary over many orders of magnitude, it is more convenient to represent them on logarithmic scale as \emph{free energies} $F_{\gamma}$. In Bayesian terminology, $\mathcal{Z}_\gamma$ is the \chg{evidence} of each coefficient set, and $F_\gamma$ is the negative \chg{log-evidence} \cite{mackay1992bayesian}. The free energy is computed directly from subsets of the precomputed empirical correlation matrix $C$ and vectors $\vec{V}_l$ of library functions with each other and with empirical derivatives (see \chg{Fig.~\ref{fig:setup} for schematic illustration and} SI for derivation).

Much like in statistical physics the free energy quantifies the balance of energy and entropy of a coarse-grained state\cite{goldenfeld}, here the free energy represents the goodness of fit of the empirical derivatives $\dot{x}_l$ with respect to all possible continuous values of coefficients within the same active set. The derivation and the final functional form of \eqref{eqn:Fgamma} are closely similar to the Akaike and Bayesian Information Criteria (AIC and BIC) that combines the likelihood of a model with a penalty based on the number of parameters \cite{akaike1974new, schwarz1978bic}. The free energy expression \eqref{eqn:Fgammaprime} selects for sparse solutions in two ways: while the AIC- and BIC-like penalty we term ``natural sparsity'' scales \chg{with} the trajectory length $n$ as $\order{1}+\order{\ln n}$ (see SI for derivation),\footnote{We thank L.~Fung and M.~Juniper for directing our attention to the natural sparsity effect which they term ``Occam factor'' in their paper \cite{fung2024rapid}.} the prior driven penalty scales as $\Lambda=\lambda\cdot n=\order{n}$. The following sections use the free energy computation to compute the full posterior and its marginals to analyze model inference in different regimes.


\section{Results}
We choose to focus our in-depth analysis \chg{in the main text} on one 3-dimensional chaotic dynamical system, the classic Lorenz attractor first integrated numerically by Ellen Fetter \cite{lorenz1963deterministic, sokol2019chaos} (Fig.~\ref{fig:free_energy}a), since it is a fairly typical chaotic system \chg{with polynomial nonlinearity \cite{gilpin2021chaos, zubov1961methods}}, and the performance of SINDy-family algorithms does not strongly correlate with different features of chaotic systems \cite{kaptanoglu2023benchmarking}. \chg{Polynomial representations cover many dynamical systems with a wide range of dynamical behaviors \cite{gilpin2021chaos, zubov1961methods}. This is in large part due to the manifestation of dominant balance physics which is often expressed as leading polynomial terms in a Taylor series expansion of more complex behavior.  Dominant balance physics is commonly used to express the mathematical similarity between highly disparate physical systems \cite{cross1993pattern}. However, there are certainly many empirical systems are not well described by polynomial forms \cite{crutchfield1987equations}. While the Z-SINDy framework, like most other SINDy forms, can handle arbitrary nonlinear library terms, we restrict this study to polynomial forms. We thus consider the library terms in form of} all monomials in variables $x,y,z$ up to 2nd order, resulting in $N=10$ library terms and $2^N=1024$ possible coefficient sets. The computational infrastructure of evaluating the library terms is based on the PySINDy package \cite{desilva2020pysindy}. \chg{Throughout this chapter we study different aspects of the Z-SINDy fit and introduce multiple graphical formats of analysis, focusing on the Lorenz system. We repeated a similar analysis flow for six other nonlinear systems, including limit cycles, decaying oscillations, a higher order nonlinear chaotic system, and a hyper chaotic system (see SI) \cite{fung2024rapid, rossler1979equation, letellier2007hyperchaos, abooee2013analysis}.}

\subsection{Free energy trends}
\begin{figure*}[t]
	\centering
	\includegraphics[width=\textwidth]{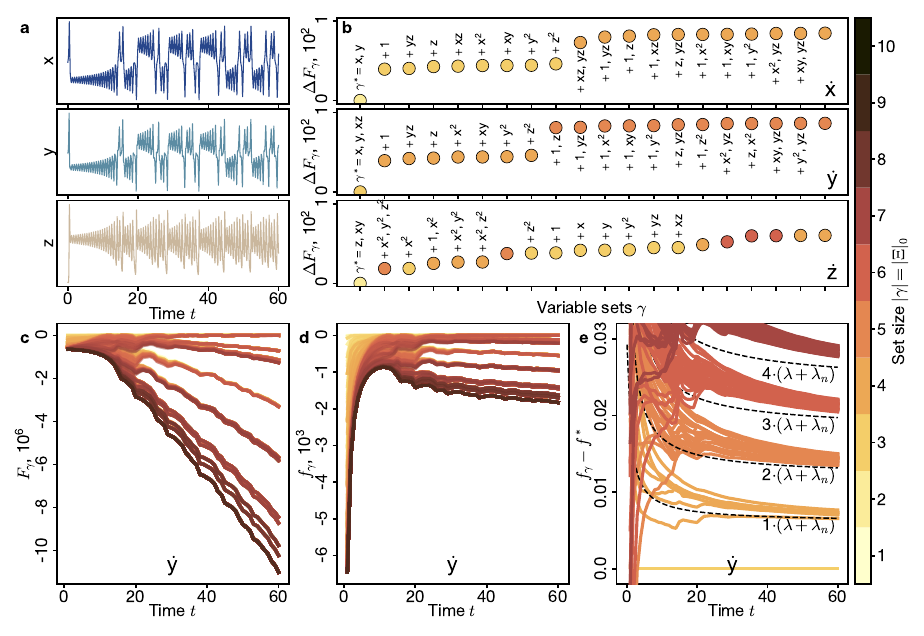}
	\caption{Free energy trends in Z-SINDy inference. (a) Sample trajectories of the Lorenz attractor sampled with time sampling period of $\Delta t=0.01$ and no noise. (b) Free energies of different coefficient sets for each dimension of dynamics relative to the best fit set $\gamma^*$ computed at $\rho=1.0$. Other coefficient sets are labeled by the extra library terms with respect to $\gamma^*$. (c) Free energy $F_\gamma$ of each of the $2^N=1024$ coefficient sets over time. (d) Free energy \emph{per data point} $f_\gamma$ of each coefficient set. (e) Relative free energy per data point $f_\gamma-f^*$ of each coefficient set, stratified by the sparsity penalty $\lambda=0.006$ and the natural sparsity penalty $\lambda_n$. The color of markers and curves corresponds to the number of coefficients in each set $|\gamma|$.}
	\label{fig:free_energy}
\end{figure*}

In order to establish intuition for free energy scaling, we use the expression \eqref{eqn:Fgamma} to compute the free energies of all variable sets for \chg{the Lorenz model}. For a constant and moderate trajectory length $T$ and sparsity penalty $\lambda$ the free energies of different variable sets form a hierarchy shown in Fig.~\ref{fig:free_energy}b for each dimension of dynamics: the correct variable set $\gamma^*$ has the lowest free energy, while the next several sets with higher free energies all have extra terms. If the trajectory data is only considered up to a variable upper limit of time $t$, the free energies of each variable set have asymptotically linear trajectories of different slopes (Fig.~\ref{fig:free_energy}c). In order to compare the slopes, we compute the \emph{intensive} free energy per data point $f_\gamma=F_\gamma /n$ that is asymptotically constant for each variable set (Fig.~\ref{fig:free_energy}d). We further disentangle the different sets by computing the intensive free energy relative to its lowest value $\Delta f_\gamma=f_\gamma-f^*$ (Fig.~\ref{fig:free_energy}e). By construction the relative free energy of the correct variable set is zero, and free energies of other sets are asymptotically stratified by the constant intensive sparsity penalty $\lambda$. While the sparsity penalty is an externally chosen hyperparameter of Z-SINDy, at low sample sizes the Bayesian inference procedure itself introduces an additional ``natural'' sparsity $\lambda_n=\ln(n/2\pi\rho^2)/2n$ that enhances the selection for sparse coefficient sets similar to the AIC and BIC (see SI for derivation).

The asymptotically constant free energy per data point in this inference problem is similar to the thermodynamic free energy of interacting particle systems. For particle systems, nonlinear scaling of free energy with system size usually implies either long-range particle interactions or strong boundary effects. Indeed, the inference free energy has a nonlinear scaling at early times ($t<10$ on Fig.~\ref{fig:free_energy}c), when the Lorenz dynamical system has only explored one lobe of the attractor and has not yet demonstrated the switching behavior. At longer times, the chaotic dynamics forget the initial condition and the effective sample size scales linearly with the amount of data, leading to a condensation of inference, as we explore in the following section.

\subsection{Inference condensation}
\begin{figure}[t]
	\centering
	\includegraphics[width=\columnwidth]{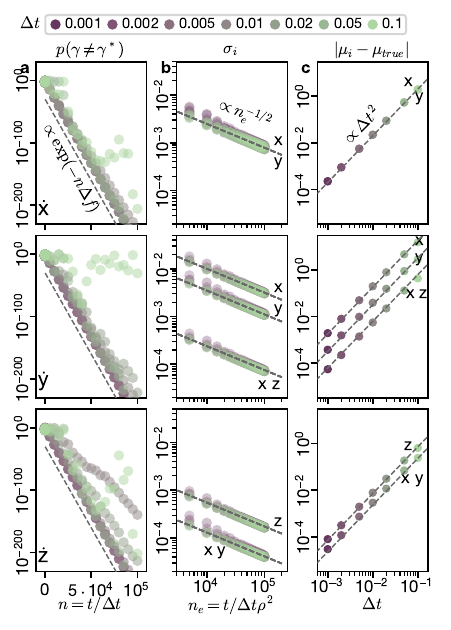}
	\caption{Condensation of Z-SINDy inference for trajectory datasets of variable runtime $t$, sampling time period $\Delta t$, resolution parameter $\rho\in\{0.1,1.0,10\}$. (a) The probability of choosing any variable set other than the one with lowest free energy decays exponentially with trajectory length. (b) The posterior standard deviation of each inferred coefficient decays as inverse square root of the trajectory length. (c) The absolute error of the posterior mean coefficient is limited by the systematic error of the finite difference derivative of the trajectory. Trend lines guide the eye to illustrate the functional form of the scalings. The line labels in (b-c) indicate which coefficient the standard deviation and the mean correspond to.
		Rows correspond to the three dimensions of dynamics. The standard deviation and error of the mean for the two variables explaining $\dot{x}$ coincide.}
	\label{fig:condensation}
\end{figure}

In order to connect the scaling of free energy to the outcomes of the inference procedure, we consider the limiting case of the posterior distribution \eqref{eqn:posteriorfact}. The probability of selecting the lowest free energy set is driven by the free energy gap between it and the next set:
\begin{align}
	p(\gamma^*)=\frac{1}{1+\sum\limits_{\gamma\neq \gamma^*} 
		e^{-(F_{\gamma}-F_{\gamma^*})}}\approx 1-e^{-n\Delta f},
\end{align}
where $\Delta f$ is the asymptotic difference of free energy per data point between the best and the second-best fitting coefficient sets. This expression implies that the probability of selecting \emph{any other} variable set decays exponentially with trajectory length across a wide range of sampling frequencies (Fig.~\ref{fig:condensation}a). While the statistical weights $\mathcal{Z_{\gamma}}$ can get exponentially large or small, risking numerical overflow or underflow problems, the values of free energies do not face that problem. The exponential suppression of sub-optimal coefficient sets implies that it is sufficient to look for the lowest free energy coefficient set at given dataset parameters and inference hyperparameters. 

Given the condensation of the discrete part of inference, what happens to the continuous part? Per \eqref{eqn:posteriorfact}, the Gaussian mixture reduces to a single multivariate Gaussian distribution with the covariance and mean parameters driven by the empirical correlations $C_{\gamma^*}$ and $\vec{V}_{l,\gamma^*}$. The posterior covariance matrix is given by $\Sigma_{\gamma^*}=\rho^2 C^{-1}_{\gamma^*}$, scaling with the resolution parameter $\rho$ but decaying with increasing trajectory length, which can be combined into an effective time scale $n_e=t/\Delta t \rho^2$. The standard deviations of the posterior along each coefficient direction have different magnitudes but identical scaling of $n_e^{-1/2}$ as in the Central Limit Theorem (Fig.~\ref{fig:condensation}b).

The posterior Gaussian mean is given by $\vec{\mu}=C^{-1}_{\gamma}\vec{V}_{l,\gamma}$, which quickly converges to a constant value, which is not necessarily equal to the ground truth $\vec{\mu}_{true}$. Since the linear regression against the nonlinear library terms aims to explain the empirical derivative, it inherits the systematic error of the numerical differentiation procedure, well known in the studies of numerical integration \cite{hu2012dispersion} but rarely highlighted in system identification. For the second-order finite difference derivative employed here the systematic error scales as $\order{\Delta t^2}$, and this scaling propagates to the error of the mean (Fig.~\ref{fig:condensation}c). For large enough trajectory length the standard deviation becomes smaller than error of the mean $\sigma_i<\abs{\mu_i-\mu_{true}}$, and thus the inference converges to a small region that does not include the ground truth values.

\chg{The scaling reported here is due to collecting an ever-longer single trajectory. An alternative way to collect more data is to stitch together many short trajectories \cite{wu2019numerical, 13schaeffer2018extracting}. Once the numerical differentiation of each trajectory is performed, the data take the form of pairs $(\dot{x},\Vec{\Theta})$, the order of which is not important. The covariances $\vec{V},C$ are computed by summation over all available data points, whether they originally came from the same continuous trajectory or different ones. We analyze the inference condensation for many short trajectories of just 10 time steps each and see qualitatively similar results, though the standard deviations $\sigma_i$ get somewhat tighter if data is collected from transient trajectories that have not yet reached the Lorenz attractor (see SI for figure and details).}

Increasing the trajectory length leads to a condensation of inference to a particular set of coefficients with a narrow range of values. However, if the amount of data asymptotically does not distinguish between inferred models, then what does? In thermodynamic systems, free energy per particle is typically a function of external thermodynamic variables, such as temperature, magnetic field, or chemical potential. Small changes of the external parameter can shift the global free energy to a different state, leading to a thermodynamic phase transition. In a similar way, small changes in the inference hyperparameter $\lambda$ or noise in the data $\eta$ can lead to \emph{abrupt} changes in the inferred model, as we show in the following sections.






\subsection{Sparsity transitions}
\begin{figure}[t]
	\centering
	\includegraphics[width=\columnwidth]{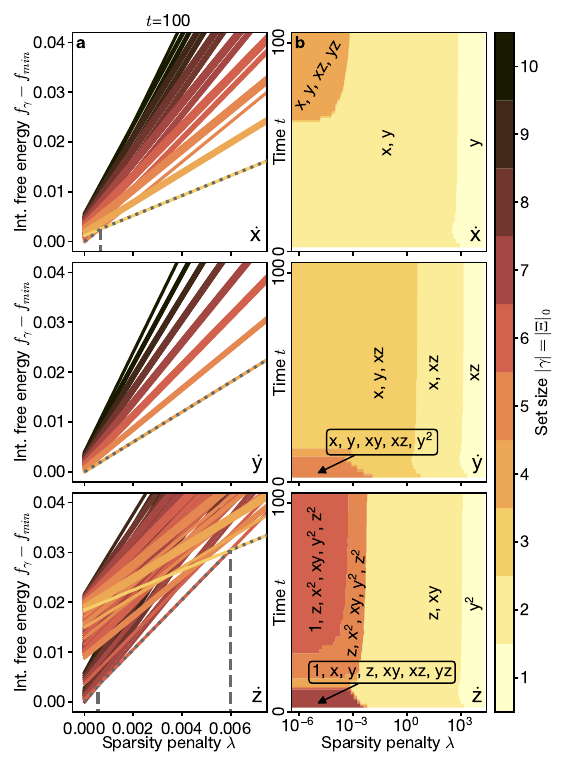}
	\caption{Sparsity penalty selects sparse solutions. (a) For a fixed trajectory of length $t=100$, the penalized intensive free energies of each coefficient set are linear functions of the sparsity penalty $\lambda$ with the slope given by the number of terms and the intercept by the goodness of fit. At any value of $\lambda$, the posterior condenses on the lowest free energy set (piecewise linear dotted lines), switching abruptly between solutions of different sparsity (vertical dashed lines). (b) As the sparsity penalty varies over many orders of magnitude, the best fit coefficient set changes abruptly several times from large to sparse, with a finite range of $\lambda$ recovering the correct set. The color of lines and backgrounds corresponds to the number of coefficient in each set.}
	\label{fig:sparsity}
\end{figure}

The goal of SINDy-family approaches is to balance the data fit with the parsimony of the inferred models, operationalized by sparsity. Instead of prescribing a particular number of equation terms, the algorithm is supposed to find it adaptively, but how exactly does the sparsity penalty parameter $\lambda$ lead to sparse solutions?

We have established in the previous section that with enough data, Z-SINDy would always select the model with lowest free energy. Given a constant dataset, the penalized free energies are linear functions of the penalty (\eqref{eqn:Fgammaprime}), with the intercept given by data fit at zero penalty, and the slope given by the number of terms. Graphically, the ensemble of all the linear functions looks like a fan plot with all-integer slopes $1,2,3,\dots$ (Fig.~\ref{fig:sparsity}a). As the external sparsity penalty $\lambda$ increases, the lowest free energy line changes in a series of abrupt transitions from lower-intercept higher-slope to higher-intercept lower-slope (shown in vertical dashed lines), similar to the plots produced by Least Angle Regression (LARS) \cite{efron2004lars}. However, because of the natural sparsity effect, the inference selects for sparse solutions even at $\lambda=0$. In order to include both external and natural sparsity, we therefore vary both $\lambda$ and the trajectory length $t$.

At moderate to high sparsity penalty $\lambda$ the selected variable sets are practically independent of trajectory length (Fig.~\ref{fig:sparsity}b), but at low $\lambda$ the selection changes in two ways. Very short trajectories do not explore the entirety of available phase space: for the Lorenz system, the trajectory stays within a single lobe of the attractor for $t<15$ (Fig.~\ref{fig:free_energy}a), and thus Z-SINDy identifies less sparse models (bottom-left corner of Fig.~\ref{fig:sparsity}b panels). For moderate trajectory length the correct coefficient set is recovered for a wide range of $\lambda$ because of the natural sparsity $\lambda_n$. For long trajectories the natural sparsity disappears, leading to identification of less sparse sets (top-left corner of Fig.~\ref{fig:sparsity}b panels). We conclude that Z-SINDy correctly identifies the sparse set of coefficients within a window of several orders of magnitude of sparsity penalty $\lambda$, but the boundaries of the region are quite abrupt. \chg{Aggregation of many short trajectories results in a similar sequence of inferred models, with larger amount of jitter for low $\lambda$ (see SI for figure and discussion).}

\subsection{Noise transition}
\begin{figure}[t]
	\centering
	\includegraphics[width=\columnwidth]{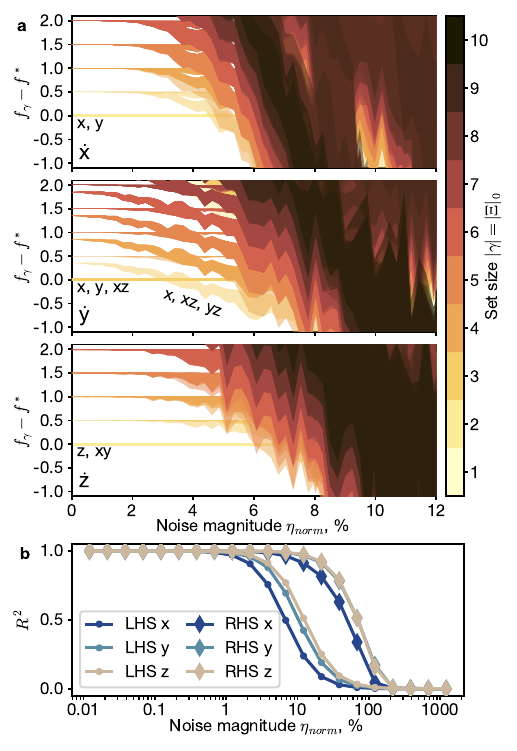}
	\caption{Additive noise prevents the identification of the correct coefficient set. (a) Each panel shows the free energies per data point with respect to the correct set for a trajectory of length $t=60$ at sampling period $\Delta t=0.01$. For each coefficient set, the shaded region shows the range of free energies across 10 realizations of random noise of given relative magnitude $\eta/\bar{\sigma}_{data}$, stratified by sparsity penalty $\lambda=0.5$. The color of the shaded regions corresponds to the number of coefficients in each set. (b) Coefficient of determination $R^2$ of the noisy left and right hand sides of the equation of motion with the correct right hand side for each dimension of dynamics, averaged over 5 noise realizations.}
	\label{fig:noise}
\end{figure}

Along with sparsity, another important limitation to the performance of SINDy algorithms is the noise in the data. The robustness of SINDy is usually measured by how fast the error in the inferred coefficients grows with noise magnitude, and thus how much noise can the inference tolerate. While the denoising approaches \cite{delahunt2022toolkit, cortiella2023priori} and the weak form SINDy \cite{messenger2021weak} improve noise tolerance significantly, they do not explain how the noise-induced breakdown happens, whether collecting a longer trajectory helps, and why denoising improves performance so much.

We seek to explain the noise-induced breakdown in free energy terms. The free energy of any variable set $\gamma$ depends not only on the trajectory length $t$, sparsity penalty $\lambda$, and noise \emph{magnitude} $\eta$, but also on the noise \emph{realization}. Within each realization, we compute the deterministic free energy per data point relative to the correct set $\gamma^*$ in each dimension, and then collect free energy statistics across multiple noise realizations. The relevant free energy statistic is not its mean but its range of fluctuations, since the inference condenses to the coefficient set with the lowest free energy.

The noise-induced transition graphically looks like an overlap between the horizontal line of the correct set and the free energy range of one of the competing sets (shaded regions in Fig.~\ref{fig:noise}a). We normalize the noise magnitude by the standard deviation of the original trajectory averaged across the dimensions $\eta_{norm}=\eta/\sigma_{data}$. At low noise magnitude $\eta_{norm}$, the correct coefficient set $\gamma^*_l$ has the lowest free energy, clearly stratified from all other ones by the sparsity penalty, and is thus selected in inference. As noise magnitude increases, at about $\eta_{norm}\approx 4\%$ the model for $\dot{y}$ changes as the set $\gamma^*=\{\textrm{x, y, xz}\}$ first overlaps with the range of free energies of $\gamma=\{\textrm{x, xz, yz}\}$ and then lies entirely above the range. Within the free energy range overlap, the inference condenses to a single coefficient set that depends on the \emph{realization} of the random noise, and thus the inference is unstable. At higher levels of noise above $\eta_{norm}\approx 6\%$ the horizontal line of the correct set lies fully above multiple overlapping free energy ranges, and thus the inference procedure would confidently select a coefficient set from many possible alternatives, none of which are correct. This inference scenario is qualitatively similar to the detection of weak communities in complex networks, where the single eigenvalue that carries community information gets buried within a continuous band of random eigenvalues \cite{nadakuditi2012graph}.

Can this inference collapse be avoided with larger amounts of data? As trajectory length $t$ increases, the width of the free energy range shrinks proportional to $t^{-1/2}$, thus reducing the range of noise magnitudes $\eta$ where the inference outcome is realization-dependent (see SI for discussion). However, since the mean free energy per data point converges to a constant $\eta$-dependent value, the takeover of the correct coefficient set $\gamma^*$ by one or more competing ones is inevitable.

What part of the data processing pipeline drives the inference collapse? SINDy approaches aim to balance the noisy versions of the left and right hand sides of \eqref{eqn:eom} $LHS_{noisy}$ and $RHS_{noisy}$ (LHS is derivatives, RHS is library terms) \emph{approximately}, but how good is that approximation? We can perform that comparison since for a synthetic dataset we have access to both clean and noisy trajectories. The clean trajectory is an exact solution to \eqref{eqn:eom} up to integration error, for which the two sides of the equation match $LHS_{true}=RHS_{true}$ and can be computed exactly by using the ground truth coefficients $\Xi_{il, true}$. We thus have three time series for each dimension, the linear correlation between which is easily measured by the coefficient of determination $R^2$.

The $R^2$ smoothly decrease from 1 at low noise to 0 at high noise, where neither LHS nor RHS of the noisy equation carry any resemblance to the truth (Fig.~\ref{fig:noise}b). While there is a slight variation of the decay point of curves between dimensions, LHS decays at almost an order of magnitude lower noise level, and is thus primarily responsible for SINDy breakdown. Since the core of SINDy is linear regression, the curves such as Fig.~\ref{fig:noise}b can be plotted for any denoising numerical derivative method and sampling time period and used as a diagnostic method to identify the limiting factor of performance and thus the most promising algorithmic improvement.

\subsection{Uncertainty Quantification}\label{sec:UQ}
\begin{figure*}[t]
	\centering
	\includegraphics[width=\textwidth]{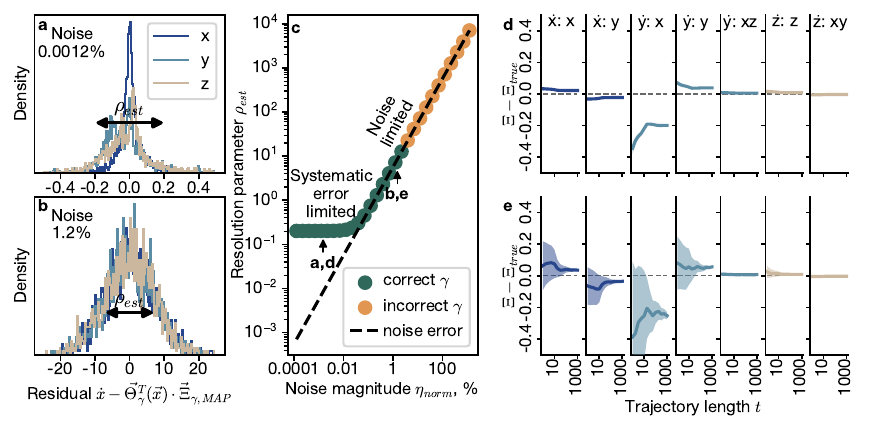}
	\caption{Uncertainty quantification from empirical residual. (a-b) Histograms of the derivative residual across the length of the trajectory for each of three dimensions of dynamics. The black arrow indicates the $\pm1\cdot\rho_{est}$ range, equal to the sample standard deviation of the residual.  (c) Inferred resolution parameter $\rho_{est}$ at each noise level. The marker color indicates whether the correct set of coefficients has been identified at given noise level. At low noise the residual is limited by the systematic error and thus the resolution parameter is constant. The dashed line indicates the analytic prediction of noise driven error in the derivative, which dominates at high noise. (d-e) Marginal posterior distributions for each of the seven active terms in the Lorenz system. The solid curves show the posterior mean relative to ground truth, the shaded region indicates $\pm 1$ standard deviation. The horizontal axes cover the range of trajectory length $t\in [1,1000]$ on logarithmic scale. Panels (a,d) are computed at low noise, panels (b,e) at moderate noise.
	}
	\label{fig:UQ}
\end{figure*}

The analysis of the noisy LHS and RHS leads to the uncertainty quantification of the inferred coefficients in the limit of low data. The posterior distributions of all coefficients are Gaussian with parameters derived from the correlations $C,\vec{V}_l$, but require knowing the resolution parameter $\rho$. The resolution parameter describes the distribution of the the residual between the noisy LHS and RHS of the dynamical equation, and can be estimated from the empirical distribution by using the Maximum A Posteriori (MAP) values of the coefficients $\vec{\mu}$:
\begin{align}
	\rho_{est}=\sqrt{\frac{1}{n\cdot d}\sum\limits_{t,l}\left( LHS_{l, noisy}-RHS_{l, noisy} \right)^2}.
\end{align}

While for each dimension the residual itself $LHS_{noisy}-RHS_{noisy}$ is expected to have a Gaussian distribution, we can check the shape of the distribution empirically (Fig.~\ref{fig:UQ}a-b). At low noise level the distribution is different for each dimension and has a complex multimodal shape with long tails (panel a), while at moderate noise the distributions for all three dimensions have a consistent Gaussian shape (panel b). As the amount of noise on the trajectory varies over several orders of magnitude, the estimated resolution parameter $\rho_{est}$ switches from a flat value driven by the systematic error of the finite sampling period to the noise driven value $\rho=\eta/\Delta t\sqrt{2}$ (Fig.~\ref{fig:UQ}c, see SI for derivation).

The resolution parameter is the last missing piece in computing the error bar on the inferred coefficients and comparing them to the ground truth. At low noise the error bar is incredibly small, so that the systematic error in the coefficients is immediately visible for all trajectory lengths (Fig.~\ref{fig:UQ}d). At moderate noise the error bar increases and overlaps with the ground truth so that the systematic error is not visible at short trajectory length. As the trajectory length increases, the error bar shrinks proportional to $t^{-1/2}$ as discussed before, resulting in the same value of the statistically significant systematic error.

\subsection{Inference phase diagram}
\begin{figure}[t]
	\centering
	\includegraphics[width=\columnwidth]{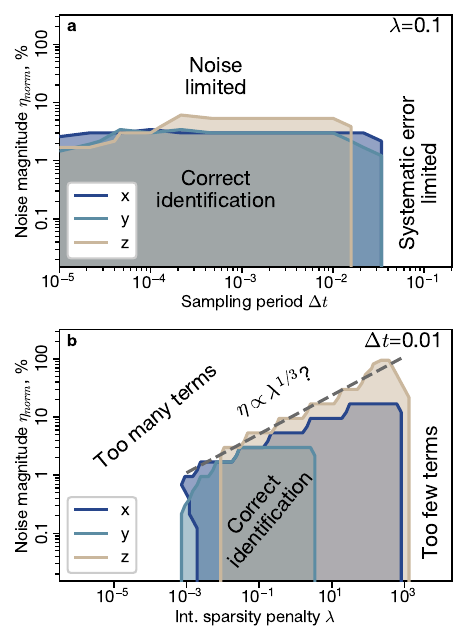}
	\caption{Inference of the correct dynamical equations is limited by the data noise $\eta$, sampling period $\Delta t$, and sparsity penalty $\lambda$. (a) Inference fails when either the noise level crosses a threshold (noise limited), or the sampling period becomes too large to resolve the smallest time scale of the dynamical system (systematic error limited), with the boundaries consistent across the three dimensions of dynamics. (b) The region of correct inference in noise and sparsity parameters is limited on three sides: too few terms at high $\lambda$, and too many terms at low $\lambda$ or high $\eta$. The dashed line guides the eye to illustrate a speculative power law on critical noise level $\eta\propto \lambda^{1/3}$.}
	\label{fig:phasediagram}
\end{figure}

Having characterized the inference breakdown with sparsity and noise separately, we can now answer the questions about the joint effect of dataset parameters and inference hyperparameters on the viability of system identification. We have shown that inference rapidly condenses to a single coefficient set with growing trajectory length $t$ that is determined by the lowest free energy \emph{per data point}. Moderate values of $t$ incur the natural sparsity effect, but it vanishes for $t\to \infty$, so that the selected coefficient set becomes independent of data quantity. The value of the resolution hyperparameter $\rho$ modulates the inference condensation and affects the size of the confidence interval for the coefficients. This value can be chosen to match the statistics of the residual $\dot{x}_l-\vec{\Theta}^T\cdot \vec{\Xi}_l$ and provide accurate uncertainty quantification.

The remaining parameters interact in a more complex way as shown in the phase diagrams of Fig.~\ref{fig:phasediagram}. The noise magnitude $\eta_{norm}$ and sampling period $\Delta t$ impose separate, orthogonal limitations (Fig.~\ref{fig:phasediagram}a). At large sampling period, the samples can no longer resolve the smallest time scales of system dynamics and result in a large \emph{systematic} error in the empirical derivative. At the same time, a small sampling period combined with presence of noise results in a large \emph{statistical} error in the empirical derivative. However, while the error of each derivative sample grows as $1/\Delta t$, the number of samples per unit time grows at precisely the same rate $1/\Delta t$ and the two effects almost cancel each other out, resulting in a nearly horizontal upper bound of correct identification (see SI for additional discussion).

The interaction of noise magnitude $\eta_{norm}$ with sparsity penalty $\lambda$ is even more complex (Fig.~\ref{fig:phasediagram}b). The noiseless regime of identification has been explored in Fig.~\ref{fig:sparsity}, revealing a finite range of several orders in $\lambda$ where the identified coefficient set has not too many and not too few terms, with a narrower range for $\dot{y}$ that has more terms in the correct coefficient set. While the sparsity penalty creates a gap between the free energies of sets of different size, growing noise level gradually reduces this gap until a free energy crossover (Fig.~\ref{fig:noise}a). Qualitatively, a larger initial gap $\lambda$ would require more noise $\eta$ to close, resulting in a positive slope of the limiting curve on the phase diagram. The trade-off has the approximate shape of a power law $\eta\propto \lambda^{1/3}$ across six orders of magnitude of $\lambda$, but the limited resolution of the phase diagram and the complexity of the free energy landscape prevent a simple explanation of this scaling, leaving an important opening for further work.


\section{Closed-loop inference}
\begin{figure*}[t]
	\centering
	\includegraphics[width=0.9\textwidth]{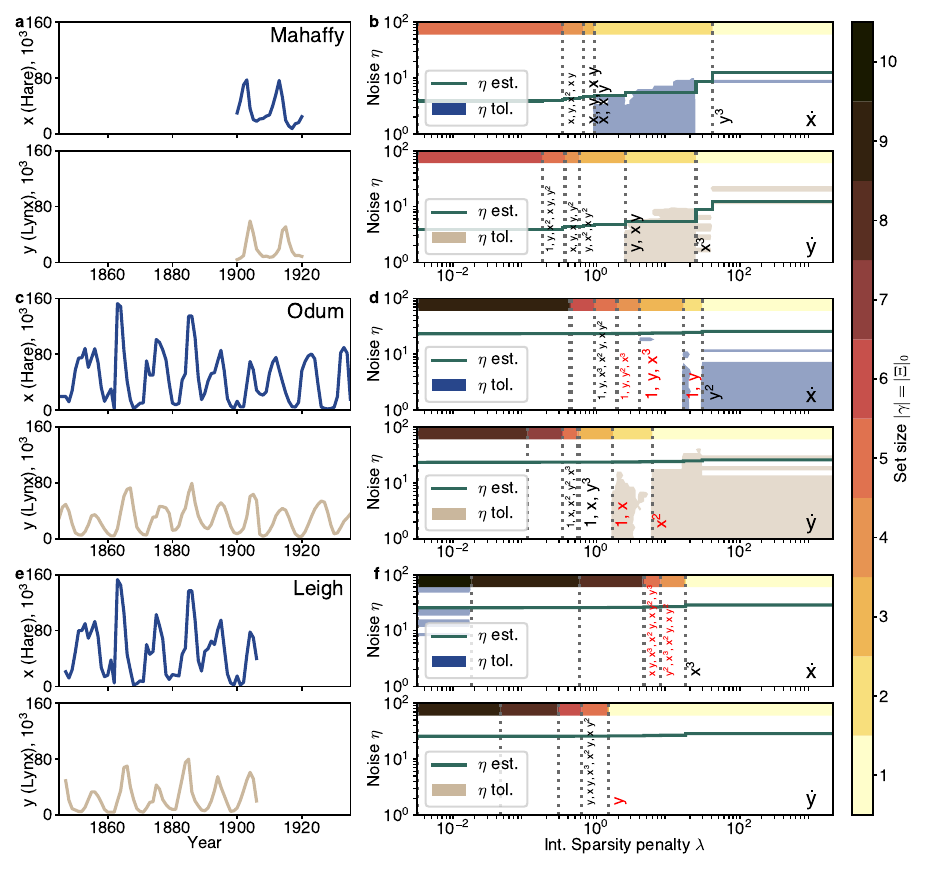}
	\caption{\chg{Closed loop inference of predator-prey interactions. (a,c,e) Raw empirical data on the number of hare and lynx pelts reported in each year within the data range for the three datasets, in units of $10^3$. (b,d,e) Noise estimate (green step curves) and tolerance diagrams (shaded regions). The vertical gray dashed lines show the sparsity-induced transitions between the models. The coefficient sets are labeled to the right of the transitions, with unstable models marked in red. The color-coded shaded stripe at the top of each panel marks the size of each coefficient set.}}
	\label{fig:closedloop}
\end{figure*}

\chg{The graphical analyses presented in the previous sections, specifically the residual distribution (Fig.~\ref{fig:UQ}a-b) and the phase diagrams (Fig.~\ref{fig:phasediagram}) are powerful diagnostic tools that quantify the limits of performance of a particular data-driven algorithm. In particular, they can be combined into a closed-loop inference system: for a given dataset one can fit and integrate a SINDy model, compute its noise tolerance, and estimate the empirical noise magnitude. If the empirical noise is above the tolerance, then the model is misleading and should be rejected.}

\chg{Here we demonstrate a proof-of-concept closed-loop analysis on the famous dataset of hare and lynx pelts collected each year by the Hudson Bay Company. The original theoretical works by Lotka \cite{lotka1925elements} and Volterra \cite{volterra1926variazioni} were \emph{inspired} by observations of fluctuations in species abundance. The resulting Lotka-Volterra equations are widely celebrated, but not directly derived from empirical data. It is important to clarify that there is not a single hare-lynx dataset but multiple versions of it. Two famous historical data compilations are the Leigh dataset (1847--1906)~\cite{leigh1968ecological} and the Odum dataset (1845--1935)~\cite{odum1971ecology}.\footnote{We use the tabulation of the Leigh and Odom datasets provided via personal communication by B.~Deng \cite{deng2018inverse}.} More recently, a version of the Hudson Bay Company dataset became widely used for pedagogical reasons in multiple courses and textbooks \cite{howard2022course, mahaffy2010lv, carpenter2018stan, wilensky2006thinking, purves1992life} and used in SINDy studies \cite{hirsh2022sparsifying, fung2024rapid}, which we refer to as the Mahaffy dataset (1900-1920) \cite{mahaffy2010lv}. While the time resolution of $\Delta t=1$ year and the qualitative shape of oscillations are consistent across the three datasets, the covered date ranges overlap only partially and the reported numerical values are different (Fig.~\ref{fig:closedloop}a,c,e).}

\chg{In order to compare the three empirical datasets, we construct the closed-loop inference as follows. First, we fit a polynomial Z-SINDy model to the entirety of each dataset while varying the the sparsity penalty $\lambda$ over several orders of magnitude, resulting in a sequence of models of different sparsity. We then estimate the variance of the residual $\rho$ following the method in Fig.~\ref{fig:UQ}a,b and convert it into an estimate of noise $\eta_{est.}=\rho\cdot \sqrt{2}$ ($\Delta t=1$) (green stepped curve in Fig.~\ref{fig:closedloop}b,d,f). We then take the maximum a posteriori model at each noise level as ground truth and evaluate the inference phase diagram for it following Fig.~\ref{fig:phasediagram}b across a range of synthetic noise magnitudes to identify the tolerance $\eta_{tol.}$ (shaded region in Fig.~\ref{fig:closedloop}b,d,f).}

\chg{We find that for the Mahaffy dataset at moderate sparsity $\lambda$ we can identify the familiar Lotka-Volterra functional form $\dot{x}\propto x+xy; \dot{y}\propto y+xy$ (coefficients omitted) with the estimated noise below the noise tolerance threshold (Fig.~\ref{fig:closedloop}b). Note that adding the second term to each of the two equations causes a significant reduction in the residual variance, but adding more terms does not change variance much. For the Odum and Leigh datasets the identified functional forms are all below noise threshold and are thus spurious identifications that are idiosyncratic to the particular noise \emph{realization} in the empirical dataset (Fig.~\ref{fig:closedloop}d,f). Addition of more terms to the model does not decrease variance significantly, and some of the identified models are unstable, defined as $x^2+y^2>10^6$ within simulation time and marked with red font on the figure.}

\chg{We thus found that the Lotka-Volterra equations are robustly recovered from the Mahaffy dataset with both our method and many other methods \cite{hirsh2022sparsifying, fung2024rapid}, explaining the popularity of this dataset in pedagogical literature and data science benchmarks. It is, however, unclear where exactly the Mahaffy dataset comes from as it is arbitrarily restricted to the narrow time range of 1900--1920 and presumably back-converted into tabular form from a graph~\cite{purves1992life}.\footnote{Carpenter \cite{carpenter2018stan} traces the origin of the numerical table as follows: ``Howard (personal communication) traces the numerical data to Joseph M. Mahaffy’s (2010) lecture notes on mathematical modeling for his San Diego State University course (Math 636) \cite{mahaffy2010lv}, which in turn cites Purves, Sadava, Orians, and Heller’s book, Life: The Science of Biology.\cite{purves1992life}'' (references ours)} The older Odum and Leigh datasets are thus not useful for learning the Lotka-Volterra equations, at least not in the differential form. These datasets have a number of other controversies such as the seemingly wrong phase order of the abundance peaks, known as the Hares Eat Lynx paradox~\cite{gilpin1973hares}. Ref.~\cite{deng2018inverse} offers a detailed history of these datasets, an extended model that accounts for the dynamics of fur trappers, and an algorithm to fit the empirical trajectories in an integral form without numerical differentiation.}


\section{Discussion}

In this paper we introduce Z-SINDy, \chg{an \emph{exact}} Bayesian version of a simple form of SINDy that we analyze through the prism of statistical mechanics to understand how it works---and, importantly, how it breaks. We explicitly separate the discrete and continuous parts of equation inference and derive a closed-form posterior probability distribution of the coefficients. The probability of a particular discrete set condenses exponentially with growing trajectory length to a single coefficient set, though the set is not guaranteed to be correct. Even if the identified library terms are qualitatively correct, the value of the coefficients is subject to both systematic and noise driven errors. A combination of natural and externally imposed sparsity penalties induce a set of discontinuous transitions from a large coefficient set to a sparse one. The inference procedure correctly identifies the sparse set when the trajectory has moderate noise but fails at large noise, where the inferred equation depends strongly not only on noise magnitude, but also its specific \emph{realization}. Before the noise induced transition, matching the residual statistics provides UQ for the inferred model. The combination of these results establishes the boundaries of applicability of SINDy, and the warning mechanisms for its breakdown. \chg{The UQ and noise tolerance can be combined to have close-form inference and diagnostic of parsimonious dynamical models.}

\chg{In more practical terms, equation learning is often framed as an inverse problem, the solutions to which are generically non-unique. Uniqueness can be mathematically assured by choosing a regularization (such as the sparsity penalty), but the regularization strength becomes a hyperparameter. A common way to treat a hyperparameter in machine learning is to ``optimize away'', usually through automated cross-validation. While cross-validation can improve model generalization, it hides the modeling choices from the user and ignores the domain-specific context of the data source. When SINDy or its versions are treated as an off-the-shelf method, the reasons for their failure or even success on a given dataset are obscure. In contrast to optimizing away, frameworks such as Z-SINDy make the exploration and discovery explicit by treating sparsity as a free parameter controlled by the modeler's subjective choices. Much like traditional modeling and problem solving is an art built on a mathematical foundation, data-driven equation learning can be seen as a modern version of that art.}




\subsection{Error analysis}
Our study is not unique in application of Bayesian inference to the SINDy framework. \chg{While previous recent work relied on costly Monte Carlo sampling from the posterior \cite{hirsh2022sparsifying, gao2022bayesian, 1wang2020perspective, 3mathpati2024discovering, 4tripura2023sparse, 6nayek2021spike} or Variational Bayes techniques \cite{3mathpati2024discovering, 5more2023bayesian, 7long2024equation}}, several studies conducted concurrently but independently from ours used the closed-form computation of model evidence \cite{niven2024dynamical, fung2024rapid}. They show that using Bayesian inference with Gaussian priors gives rise to the effect we term ``natural sparsity'', but do not consider an explicitly sparsity-promoting prior such as Bernoulli-Gaussian. \chg{As a result, they focus either on the probability of identifying the correct solution form \cite{fung2024rapid} or the $L_2$ norm of the residual \cite{niven2024dynamical}  at constant noise and trajectory length.} \chg{In contrast, we} provide both the boundaries of correct identification \chg{of the model form}, as well as the asymptotic scaling of coefficient errors \chg{and uncertainties} with dataset size and sampling period.

\chg{Early system identification papers acknowledged the error in the determined coefficients \cite{19regalia1994unbiased, 15rudy2019data}, but were unable to separate the statistical (noise-induced) from systematic (differentiation-induced) errors \cite{rudy2017pde}. Early forms of system identification focused on the Pad\'e approximants traced the error in the coefficient to noise in both the matrix and vector components of regression ($C$ and $\vec{V}$ in our notation) \cite{19regalia1994unbiased}. Our study proposes a simple visual diagnostic of the origins of error in Fig.~\ref{fig:noise}b, which shows that for many practical cases the error in the vector due to the derivative estimation $\dot{\vec{x}}$) is vastly higher than the error in the matrix due to evaluation of nonlinear library terms $\Theta(\vec{x})$ at noisy arguments. At the same time, the derivative estimation is contaminated by both bias and variance, thus leading to both systematic and statistical error (see SI for derivation).}

\chg{The different origin and scaling of the two error terms, as well as their mitigation were only recognized recently in both strong and weak forms of SINDy \cite{18tang2023weakident}. The numerical differentiation error can typically be reduced by using higher-order schemes \cite{17chen2023data}, whereas noise is most commonly treated with the weak form of SINDy \cite{messenger2021weak, 10messenger2022asymptotic}. At the same time, the weak formulation acts as an effective low-pass filter, thus distorting the high-frequency components of the time series that might be important to the dynamics \cite{fung2024rapid}. Ref.~\cite{20bortz2023direct} assumes that the model form is known and focuses on \emph{estimation} rather than \emph{identification} of dynamics and provides the most comprehensive error analysis for the weak form, and thus can in the future serve as a basis for a weak formulation of an exact Bayesian method. Whereas all of these studies were focused on the scaling and bounding of a point estimator error, the work presented here is Bayesian, thus expressing the statistical error in the scaling of the error bar width, and the systematic error in the offset between the center of the error bar and the ground truth (Fig.~\ref{fig:UQ}). In case of the Lorenz system presented in the main text, the systematic error in coefficients inferred from a long trajectory is roughly equivalent in low-noise and moderate-noise cases (Fig.~\ref{fig:UQ}d vs e), whereas for other systems addition of noise can induce a significant systematic error (see SI). The systematic errors depend on the subtle interplay between the sampling time step $\Delta t$, the noise magnitude $\eta$, and the natural timescales of the studied dynamical system, which opens avenues for further investigation.}

\subsection{Statistical mechanics analogies}
Instead of searching for the best fit alone, we characterize the whole landscape of models on a logarithmic scale by computing their free energies. This computation integrates out the continuous coefficient values and focuses on the coefficient sets, building a connection with coarse-grained statistical mechanics and thermodynamics of interacting particle systems \cite{goldenfeld}. The consideration of long time trajectories is akin to the thermodynamic limit and motivates the evaluation of intensive free energy per data point similar to the free energy per particle. The thermodynamic cost of adding another ``particle'' type into the model, i.e. another SINDy term, is given by the sparsity penalty $\lambda$ that functions as (negative) chemical potential. The chemical potential of all ``particles'' is identical, but the interactions between them are driven by the trajectory data and thus select a particular ``particle'' set, i.e. a set of active SINDy library terms. As the parameters of the dataset or the hyperparameters of the inference procedure are adjusted, the inferred model can change discontinuously, akin to a phase transition or a dynamical system bifurcation. The pattern of free energies of different coefficient sets exhibits rich structure, opening avenues for further study, including connection to the entropy metrics of dynamical system trajectories \cite{gaspard1993noise}.

The connections between statistical mechanics, statistical inference, and machine learning have a rich history, focusing primarily on the average-case behavior of prediction risk \cite{langley1999tractable}. Statistical mechanics helped identify and describe multiple inference regimes and phase transitions between them, from network structure inference to constraint satisfaction to compressed sensing \cite{ganguli2010statistical, nadakuditi2012graph, zdeborova2016statistical, krzakala2022notes}. The so-called replica method has been used to characterize the regularized least squares regression that also lies at the core of SINDy \cite{bereyhi2020statistical}, identifying the regimes where local greedy algorithms can efficiently identify the optimal set of predictor variables \cite{obuchi2018statistical}. However, statistical physics studies often consider the large-data case $n\to\infty$ in which, in our notation, $\abs{\gamma}/n=const$, i.e. the set of selected coefficients is sparse but growing. This paper enriches the discussion by analyzing the case of dynamical systems in which the number of differential equation terms staying constant regardless of trajectory length $\abs{\gamma}=const$, paying attention to the specific interpretable nature of individual terms, and painting a detailed picture of inference breakdown.

\subsection{Integration with other SINDy techniques}
The system identification scenario considered here aims to extract a parsimonious model \cite{kutz2022parsimony}, but focuses only on one aspect of parsimony---sparsity---over other aspects such as discovery of coordinates and parametric dependencies, both of which have been included in other data-driven methods. The coordinate discovery has been addressed by combining dynamics discovery with an autoencoder neural network that automatically discovers the sparse coordinates either in the optimization framework \cite{champion2019data, bakarji2023discovering} or the Bayesian framework \cite{gao2022bayesian}. The parametric dependence can be inferred by including a parameter library along with the dynamical equation library \cite{nicolaou2023parameter}. The parsimony requirements can be supplemented by other desired features of dynamical systems such as global stability \cite{kaptanoglu2021stability, peng2024local}. Integrating fast posterior computations from the present paper with nonlinear coordinate transformation discovery, parametric inference, and dynamic stability remain important avenues for further work. Within the analysis of trajectories, additional improvements can be achieved by a finer-scale analysis of free energy fluctuations with respect to the general trend $F\sim f\cdot n$, as well as active learning to proactively sample the unexplored parts of phase space akin to the technique suggested in Refs.~\cite{fasel2022ensemble, fung2024rapid}.

All SINDy approaches rely on regression and thus require a reliable estimation of the dependent variable, the trajectory derivative. While here we employ the simplest derivative method, finite difference, most other contemporary SINDy algorithms rely on some version of denoising derivative, such as total variation \cite{chartrand2011numerical}, spectral derivative \cite{schafer2011savitzky}, weak form \cite{messenger2021weak}, or basis expansion \cite{north2022bayesian, hokanson2023simultaneous}, which drastically improve the benchmark noise tolerance. However, such performance improvement might be misleading in quantifying the parameter uncertainty since the data uncertainty is discarded \cite{north2022bayesian}. Moreover, the denoising process is itself parametric and trades random noise for a systematic error in the derivative and the library terms \cite{chartrand2011numerical, hu2012dispersion}, requiring hyperparameter optimization \cite{vanbreugel2020numerical, vanbreugel2022pynumdiff}. 
\chg{The weak formulation of SINDy \cite{messenger2021weak, 10messenger2022asymptotic} is a promising direction for improving noise tolerance, has been used in the ensemble resampling version of SINDy \cite{fasel2022ensemble}, and can be incorporated into the exact Bayesian framework of Z-SINDy in the future. However, the weak formulation can act as an effective low-pass filter, suppressing the rapid dynamics that might be important \cite{fung2024rapid}.}
For an experimental system where the ground truth equations are unknown, it is thus unclear whether small variations of the dynamical variables are due to measurement noise or genuine fine-scale dynamics. Z-SINDy does not make a choice between those options by keeping $\rho$ as a free parameter of how closely the linear combination of library terms should approximate the derivative. Since both the derivative and the library terms suffer from noise contamination, it is challenging to pick $\rho$ \emph{a priori} but it can be estimated after the SINDy fit from the remaining unexplained variance in the derivative $\dot{x}-\Theta^T\cdot \Xi$ (see \ref{sec:UQ} and SI for additional discussion).

The free energy analysis presented here makes a prediction of the identification phase diagram for a known ground truth dynamical model. On one side, the residual distribution (Fig.~\ref{fig:UQ}a-b), the phase diagrams (Fig.~\ref{fig:phasediagram}), and the $R^2$ plots (Fig.~\ref{fig:noise}b) are powerful diagnostic tools that quantify the limits of performance of a particular numerical algorithm and thus suggest how the boundary of detectability can be pushed, with the most immediate gains available through denoising. On the other side, the analysis establishes the noise tolerance that can be part of a closed-loop inference system: for a given dataset one can fit and integrate a SINDy model, compute its noise tolerance, and estimate the empirical noise magnitude \chg{(Fig.~\ref{fig:closedloop})}.

\subsection{Computational considerations}
The main computational advantage of Z-SINDy is the closed form evaluation of the posterior distribution \chg{within a few seconds on a laptop computer}, avoiding the costly techniques of bootstrap resampling \cite{fasel2022ensemble}, Markov Chain Monte Carlo \cite{gao2022bayesian}, and repeated ODE integration \cite{hirsh2022sparsifying} that are required in other UQ system identification methods \chg{(see SI for a detailed comparison table)}. On the other side, an important computational limitation to our method is the combinatorial enumeration of coefficient sets. While the free energy \eqref{eqn:Fgamma} of any particular set $\gamma$ is very cheap to compute, the number of all possible sets grows exponentially with the number of library terms $\abs{\{\gamma\}}=2^N$\chg{; in other words, Z-SINDy has aggressive scaling with the number of library terms but a very small prefactor that makes enumeration still viable in many practical cases}.

In this paper we chose to evaluate free energies exhaustively for all sets to illustrate their scaling behavior. At the same time, the free energies of different sets are heavily stratified (e.g. Fig.~\ref{fig:free_energy}c), and only a small fraction of sets are involved in sparsity- or noise-induced breakdowns. In order to make Z-SINDy more computationally efficient and thus tractable for realistic systems, future work should aim to understand the patterns of free energies better to reduce the number of sets to evaluate, for instance by using greedy methods \cite{fung2024rapid}, subset selection \cite{foster1994risk}, least angle regression \cite{efron2004lars}, mixed-integer optimization \cite{bertsimas2023learning}, or branch-and-bound methods \cite{bertsimas2020sparse}. \chg{Other avenues for computational speedup include using multiple threads to evaluate the embarrassingly parallel set of free energies, taking advantage of the boundary method to iteratively compute the inversions of progressively larger matrix inverses $C_\gamma^{-1}$, or using Monte Carlo sampling of sets.} At the same time, we hope that the physical intuition of coarse-graining, chemical potentials, and free energies per data point would inform the further development of statistical methods for dynamical systems. 

Python code for our implementation of Z-SINDy is available at \url{https://github.com/josephbakarji/zsindy}.

\section*{Acknowledgments}
The authors would like to thank U.~Fasel, L.~Fung, L.~M.~Gao, M.~Juniper, P.~Langley, J.~Michel\chg{, R. Roy, and S.~Still} for helpful discussions\chg{, B.~Deng for providing the tabulated predator-prey data from Ref.~\cite{deng2018inverse}, anonymous referees for extensive reference suggestions,} and L.~D.~Lederer for administrative support. This work uses Scientific Color Maps for visualization \cite{crameri2023color}. The authors acknowledge support from the National Science Foundation AI Institute in Dynamic Systems (grant number 2112085). J.~B. acknowledges the support of the Center for Advanced Mathematical Sciences, and the Artificial Intelligence, Data Science and Computing Hub at the American University of Beirut.

\bibliography{biblio}
	
\end{document}


\title{Statistical Mechanics of Dynamical System Identification:\\Supplementary Information}
	\author{Andrei A.~Klishin}
        \email{aklishin@uw.edu}
	\author{Joseph Bakarji}
	\author{J.~Nathan Kutz}
	\author{Krithika Manohar}
		\email{kmanohar@uw.edu}

	\date{\today}

	\maketitle

\section{Statistical mechanics derivation}
\subsection{Prior distribution}
In order to perform inference, the Bayesian prior needs to be choosen in a functional form that promotes sparsity. Usual choices of the prior also require it to be differentiable, but we relax that condition in order to make the separation of discrete and continuous degrees of freedom explicit. The qualitative shape of the prior is "spike-and-slab", i.e. some weight concentrated near zero and the rest spread across a wide range, but there are different ways to express it mathematically.

We choose an expression of the following form:
\begin{align}
	p(\Xi)=\prod\limits_i \frac{1}{1+e^{-\lambda}} \left( \delta(\Xi_i)+e^{-\Lambda} w(\Xi_i) \right),
	\label{eqn:prior}
\end{align}
where $\delta(\cdot)$ is the Dirac delta function and $w(\cdot)$ is in principle any wide distribution on the real line. In case $w(\cdot)$ is Gaussian, this prior choice is known as Bernoulli-Gaussian and has been used in other studies of $l_0$ regularized least squares \cite{obuchi2018statistical}.

The expression \eqref{eqn:prior} is symmetric under exchange of coefficient labels, since there is no preference towards any particular set of coefficients before any data is seen, other than sparsity. This intuition can be made even more explicit by performing the product and opening all the brackets. Since there are $N$ multiplicative terms in the product and each includes 2 unique additive terms, the resulting expression is a sum over $2^N$ possible sets of terms $\gamma$ (power set of the coefficients):
\begin{align}
	p(\Xi)=\frac{1}{(1+e^{-\Lambda})^N} \sum\limits_\gamma \left( \prod\limits_{i\notin \gamma} \delta(\Xi_i) \right) \left( \prod\limits_{i\in \gamma} w(\Xi_i) \right) e^{-\Lambda \abs{\gamma}},
	\label{eqn:priorfactor}
\end{align}
where the size of the set $\abs{\gamma}$ equals to the number of nonzero coefficients, i.e. the $l_0$ pseudonorm $\abs{\Xi}_0$.

\subsection{Free energy calculation}
By combining the prior (\eqref{eqn:prior}) with the forward model (Eq.~3 of the main text), we can write down the posterior distribution:
\begin{align}
	p(\Xi|x)=\frac{p(\Xi)p(x|\Xi)}{\int_\Xi p(\Xi)p(x|\Xi)}=\frac{\sum\limits_\gamma \mathcal{Z}_\gamma p(\Xi_\gamma|x,\gamma)}{\sum\limits_\gamma \mathcal{Z}_\gamma},
	\label{eqn:posterior}
\end{align}
where both the numerator and the denominator break into a sum over all sets $\gamma$: the numerator sums probability distributions, while the denominator sums scalar factors. Note that every single term in both the numerator and the denominator has the data-independent normalization factors $\frac{1}{(1+e^{-\Lambda})^N}$ (from the prior) and $(2\pi \rho^2)^{-N/2}$ (from the forward model) which cancel out. We also note that the posterior probability factorizes over the dimensions of dynamics $x_l$, thus the following computations can be performed for each dimension separately.

We start with considering the denominator, where each term has the following shape:
\begin{align}
	\mathcal{Z}_\gamma(s)=\int\limits_{\Xi_\gamma} d\Xi_\gamma \exp(-\frac{1}{2\rho^2}\sum\limits_t (\dot{x}-\Theta^T_\gamma (\vec{x})\cdot \vec{\Xi}_\gamma)^2) \left( \prod\limits_{i\in \gamma} w(\Xi_i) \right) e^{-\Lambda \abs{\gamma}},
\end{align}
where the subscript $\gamma$ denotes that only the vector elements within a particular set are used, and $s$ is a prior hyperparameter. In order to evaluate this integral, we adopt a particular form of Gaussian prior $w(\cdot)$ of width $s$ as well as the fact that $\Lambda$ is a free parameter that can be redefined to cancel out the Gaussian normalization. In particular, we take:
\begin{align}
	w(v)&=\frac{1}{\sqrt{2\pi s^2}}\exp(-\frac{v^2}{2 s^2})\\
	\Lambda&\to \Lambda-\ln(\sqrt{2\pi s^2}).
\end{align}

We now compute the resulting Gaussian integral in the limit of an infinitely wide Gaussian prior $s\to \infty$:
\begin{align}
	\mathcal{Z}_\gamma=&\lim\limits_{s\to\infty} \mathcal{Z}_\gamma(s)=\lim\limits_{s\to\infty} \int\limits_{\Xi_\gamma} d\Xi_\gamma \exp(-\frac{1}{2\rho^2}\sum\limits_t (\dot{x}-\Theta^T_\gamma (\vec{x})\cdot \vec{\Xi}_\gamma)^2 - \frac{\vec{\Xi}^T_\gamma \cdot \vec{\Xi}_\gamma}{2s^2}) e^{-\Lambda \abs{\gamma}}\\
	=&\lim\limits_{s\to\infty} \int\limits_{\Xi_\gamma} d\Xi_\gamma \exp( -\frac{1}{2\rho^2}\sum\limits_t \dot{x}^2 + \frac{\vec{V}^T_\gamma}{\rho^2} \cdot \vec{\Xi}_\gamma - \frac{1}{2}\vec{\Xi}^T_\gamma \left( \frac{C_\gamma}{\rho^2}+\frac{1}{s^2} \right) \vec{\Xi}_\gamma ) e^{-\Lambda \abs{\gamma}},
	\label{eqn:Zintegral}
\end{align}
where we defined the library-derivative and library-library covariances as follows:
\begin{align}
	\vec{V}\equiv & \sum\limits_t \dot{x}\vec{\Theta}(\vec{x}) \label{eqn:V}\\
	C\equiv & \sum\limits_t \vec{\Theta}(\vec{x}) \vec{\Theta}^T(\vec{x}), \label{eqn:C}
\end{align}
which can be precomputed once for all $N$ library terms and then subsampled for any particular set $\gamma$.

The resulting integral \eqref{eqn:Zintegral} is a standard multivariate Gaussian integral that has a closed-form solution. The limit $s\to\infty$ converges if the matrix $C_\gamma$ is invertible, thus requiring the library functions to be non-collinear when evaluated on the trajectory. This condition can be achieved for a suitable choice of library functions and in particular holds numerically for all coefficient sets $\gamma$ in the present paper. Note that the limit $s\to \infty$ also requires an infinite adjustment in the value of the sparsity penalty $\Lambda$. This adjustment is inevitable as the statistical weight integrals are performed over domains of different support dimension, e.g. 3D vs 4D. The closed-form computation of the statistical weights is only possible when the limit is taken along a particular curve in the $(s,\Lambda)$ subspace of hyperparameters, similar to Cauchy principal value of integrals.

The closed-form expression for the integral (the statistical weight of set $\gamma$) can be thus written as follows:
\begin{align}
	\mathcal{Z}_\gamma=\mathcal{Z}_0\sqrt{\frac{(2\pi\rho^2)^{\abs{\gamma}}}{\det C_\gamma}} \exp(\frac{1}{2\rho^2} \vec{V}^T_\gamma C^{-1}_\gamma \vec{V}_\gamma) e^{-\Lambda \abs{\gamma}},
	\label{eqn:Zgamma}
\end{align}
where $\mathcal{Z}_0$ is the statistical weight of an empty set of coefficients $\gamma=\varnothing$, closely related to the variance of the derivative time series. Since exponential expressions are both not convenient to deal with and risk underflows and overflows in numerical conversations, we instead consider the statistical weight on logarithmic scale as \emph{free energy} of each coefficient set:
\begin{align}
	\mathcal{Z}_\gamma \equiv& e^{-F_\gamma} = e^{-F'_\gamma} e^{-\Lambda \abs{\gamma}}\\
	F'_\gamma =& -\ln \mathcal{Z}_0 +\frac{1}{2}\ln(\det C_\gamma)-\frac{\abs{\gamma}}{2}\ln(2\pi \rho^2)- \frac{1}{2\rho^2} \vec{V}^T_\gamma C^{-1}_\gamma \vec{V}_\gamma,
	\label{eqn:Fgamma}
\end{align}
as presented in the main text.

\subsection{Natural sparsity effect}
The natural sparsity effect appears from the consideration of the middle two terms of the free energy \eqref{eqn:Fgamma}. The first term corresponds to the variance of the derivative time series:
\begin{align}
	-\ln \mathcal{Z}_0 = \frac{1}{2 \rho^2} \sum\limits_t \dot{x}^2,
\end{align}
whereas the last term corresponds to the variance explained by the best fit linear regression against the library terms from the chosen set.

The remaining two terms appear from the consideration of coefficient values around the best-fit value. In order to interpret them in terms of natural sparsity, we consider how free energy scales with dataset size $n$. For trajectories that explore a finite attractor for long enough, the new data points of $x$ and $\dot{x}$ are sampled from the same distribution, and thus the covariances \eqref{eqn:V}, \eqref{eqn:C} should grow proportionally to the sample size:
\begin{align}
	\vec{V}_l\propto n;&\quad C\propto n\\
	-\ln \mathcal{Z}_0 \propto n;&\quad -\frac{1}{2\rho^2} \vec{V}^T_\gamma C^{-1}_\gamma \vec{V}_\gamma \propto n,
\end{align}
which motivates considering the intensive free energies per data point $f_\gamma \equiv F_\gamma /n$. The scaling of the two middle terms is sublinear in $n$, or more precisely,
\begin{align}
	\frac{1}{n}\left( \frac{1}{2}\ln(\det C_\gamma)-\frac{\abs{\gamma}}{2}\ln(2\pi \rho^2) \right)=
	\frac{1}{n}\left( \frac{1}{2}(\abs{\gamma}\ln n+\ln \det C_\gamma^*)-\frac{\abs{\gamma}}{2}\ln(2\pi \rho^2) \right) = \abs{\gamma}\frac{\ln(n/2\pi\rho^2)}{2n}+\frac{\ln \det C_\gamma^*}{n},
\end{align}
where we used the identity that for a $k\times k$ matrix $A$ and a scalar number $a$, $\det(aA)=a^k\det(A)$. The first term depends on the number of terms in the set $\gamma$, but not on their identity or the data, thus constituting the natural sparsity effect. The second term depends on the data via the covariance matrix per data point $C^*\equiv C/n$ and thus does not have an intuitive explanation. However, since the two terms decay at rates $\order{\ln n/n}$ and $\order{1/n}$, respectively, the natural sparsity effect is likely to be more important.

\chg{The origins of the natural sparsity term can be interpreted from several points of view.}

\chg{\emph{Laplace method of integration} approximates multivariate integrals around the global maximum of the integrand as $\textsf{magnitude}\times \textsf{hypervolume}$. Here the \textsf{magnitude} scales exponentially with the amount of data, while the hypervolume scales as a power law $\textsf{width}^\abs{\gamma}$. Once we consider the value of the integral on log scale, $\ln \textsf{magnitude}$ gives the leading-order behavior and $\abs{\gamma}\ln \textsf{width}$ the subleading behavior.}

\chg{\emph{Statistical physics} considers the fluctuations of the fit around the global energy minimum (loss function minimum). Since the energy grows only quadratically with deviations from the global minimum, it allows for small fluctuations, resulting in an entropy term that scales with the local dimension of the energy minimum $\abs{\gamma}$ but only logarithmically with system size $n$. The sum of energy and entropy components results in the \emph{free energy} of a given coefficient set $\gamma$.}

\chg{\emph{Statistical inference} aims to compare models of different complexity near the respective maximum likelihood points. By using the quadratic approximation of local log-likelihood, one can derive the penalty that depends on the number of parameters but not their identity, known as \emph{Bayesian information criterion} \cite{schwarz1978bic} Since the likelihood scales linearly with the amount of data and BIC only logarithmically, it would allow for more complex models to fit larger data. In case of inferring dynamical equations, we desire a constant model sparsity for any trajectory length, and thus have to impose additional explicit sparsity via the Bernoulli-Gaussian prior.}

\subsection{Noisy free energy}
In case noise is present in the dataset, the covariances $C$ and $\vec{V}_l$ become functions of the noise realization, resulting in the ranges of free energies shown in Fig.~5a of the main text. Nevertheless, we can make an argument about the asymptotic scaling of the noise.

If the overall length of the trajectory is $t$ and it explores a finite region of phase space, we can divide it into uncorrelated ``chunks'' of identical length $\tau$ so that the whole trajectory consists of $t/\tau$ chunks. The covariances can then be broken down into the deterministic and the stochastic parts:
\begin{align}
	C=C^\tau \frac{t}{\tau}+\sum\limits_{i=1}^{t/\tau} \Delta C^\tau_i;\quad
	\vec{V}_l=\vec{V}^\tau_l \frac{t}{\tau}+\sum\limits_{i=1}^{t/\tau} \Delta \vec{V}^\tau_{l,i},
	\label{eqn:detstoch}
\end{align}
where $\Delta C^\tau_i$ and $\vec{V}^\tau_{l,i}$ are identically distributed for all $i$ and zero-mean since the mean part can be moved into the deterministic term. The dependence of $C^\tau, \Delta C^\tau, \vec{V}^\tau_l, \Delta \vec{V}^\tau_l$ on the noise magnitude $\eta$ is complex, but we can get approximate results if $\eta$ is fixed and only $t$ varies.

For long trajectories the set-dependent part of free energy \eqref{eqn:Fgamma} is dominated by the last term. We plug in the covariance expressions \eqref{eqn:detstoch}, open the brackets, and note that the stochastic parts of the covariances $\Delta C^\tau, \Delta \vec{V}^\tau_l$ can only correlate with themselves and each other within the same chunk. While the resulting expression of free energy is complex, it would consist of $t/\tau$ deterministic terms $F^\tau_\gamma$ and $t/\tau$ stochastic terms $\Delta F^\tau_\gamma$. The variance of free energy across noise realizations thus grows proportional to the number of chunks $t/\tau$ and thus the trajectory length $t$. When we compute the free energy per data point $f_\gamma=F_\gamma/n=F_\gamma*\Delta t/t$, the resulting standard deviation scales as $t^{-1/2}$, while the deterministic part remains constant. As a result, the free energy ranges of alternative coefficient sets in Fig.~5a of main text would shrink but keep the same mean.

\subsection{Inclusion probability}
The posterior probability distribution \eqref{eqn:posterior} is defined for the continuous values of the coefficients, but can also inform the inference on the discrete level of sets $\gamma$. The probability of choosing any particular set $\gamma$ is particularly simple:
\begin{align}
	p(\gamma|x)=\mathcal{Z}_\gamma / \sum\limits_\gamma \mathcal{Z}_\gamma,
\end{align}
whereas the probability of inclusion of any particular variable can be found by summing over all sets which contain the variable:
\begin{align}
	p(\Xi_i\neq 0|x)=\frac{\sum\limits_{\gamma \supset i} \mathcal{Z}_\gamma }{\sum\limits_\gamma \mathcal{Z}_\gamma}.
\end{align}

In the main text we show that for a large enough dataset the sum over the statistical weights of all variable sets is exponentially dominated by its largest term, corresponding to the set with lowest free energy per data point $f_\gamma$. In this case the probability $p(\gamma|x)$ condenses to a single set.

\subsection{Posterior distribution}
The posterior distribution \eqref{eqn:posterior} is a Gaussian mixture, but it reduces to a single multivariate Gaussian once inference condenses to a single coefficient set $\gamma$. It is thus useful to transform the expression into a standard Gaussian form for any set $\gamma$, whether it condensed or not. The direct form from the posterior is:
\begin{align}
	p(\Xi_\gamma | \gamma)=\frac{1}{\mathcal{Z}_\gamma} \exp(-\frac{1}{2\rho^2}\sum\limits_t 
	\left( \dot{x}-\vec{\Theta}_\gamma^T \vec{\Xi}_\gamma \right)^2),
\end{align}
where we have previously computed the normalization $\mathcal{Z}_\gamma$ in Eqn.~\ref{eqn:Zgamma}. Now, however, we need to transform the above expression to the standard multivariate Gaussian form:
\begin{align}
	p(\Xi_\gamma | \gamma)=(2\pi)^{-\abs{\gamma}/2}\det(\Sigma_\gamma)^{-1/2}
	\cdot\exp(-\frac{1}{2}(\vec{\Xi}_\gamma -\vec{\mu}_\gamma)^T \Sigma_\gamma^{-1} (\vec{\Xi}_\gamma -\vec{\mu}_\gamma)).
\end{align}

By opening the brackets and matching terms, we get the following expressions for the multivariate mean vector and covariance matrix:
\begin{align}
	\vec{\mu}_\gamma=&C_\gamma^{-1}\vec{V}_\gamma \label{eqn:mu}\\
	\Sigma_\gamma=&\rho^2 C_\gamma^{-1},
\end{align}
which allow a particularly simple way to marginalize the distribution. Since the joint distribution over all variables $\vec{\Theta}_\gamma$ is a multivariate Gaussian, the marginal distribution over one variable $i$ from the set $\gamma$ is a univariate Gaussian:
\begin{align}
	p(\Xi_i | \gamma)=&\begin{cases}
		\frac{1}{\sqrt{2\pi \sigma^2_{i|\gamma}}} \exp(-\frac{(\Xi_i-\mu_{i|\gamma})^2}{2\sigma^2_{i|\gamma}}) &,i\subset \gamma \\
		\delta(0) &, \textsf{otherwise}
	\end{cases}\\
	\mu_{i|\gamma}=&(C_\gamma^{-1}\vec{V}_\gamma)_i \\
	\sigma_{i|\gamma}^2=&\rho^2 (C_\gamma^{-1})_{ii},
\end{align}
where importantly the vector and matrix operations need to be performed before reading off specific indices. Since a generic library cross-covariance matrix is not diagonal, and the diagonal of the inverse is not equal to the inverse of the diagonal:
\begin{align}
	(C_\gamma^{-1})_{ii}\neq (C_{\gamma,ii})^{-1},
\end{align}
the marginal distribution over variable $i$ depends on which other variables are included with it in the set $\gamma$. The functional form remains Gaussian, but the mean and variance vary.

The marginal distribution of any variable from the set $\gamma$ is thus either a Gaussian or a delta function. We can then get the full marginal by summing over all possible variable set:
\begin{align}
	p(\Xi_i)=\sum\limits_{\gamma} p(\Xi_i|\gamma)P(\gamma)=\sum\limits_{\gamma} p(\Xi_i|\gamma) \frac{\mathcal{Z}_\gamma}{\mathcal{Z}},
	\label{eqn:marginal}
\end{align}
resulting in a Gaussian mixture. For low-noise scenarios when one variable set clearly dominates, the Gaussian mixture is dominated by one term, and thus the marginal would look like a simple univariate Gaussian. The standard deviation of the Gaussian is directly proportional to the resolution parameter $\rho$.

\section{Uncertainty quantification}
\subsection{Fitting $\rho$}

In Fig.~3 of main text we show that the choice of the resolution parameter $\rho$ that goes into free energy calculations does not qualitatively change the selected coefficient set or the \emph{mean} values of the coefficients and merely rescales the convergence rate. However, $\rho$ also multiplies the posterior covariance matrix and thus the posterior standard deviations on the parameters. Knowing $\rho$ is thus essential in order to provide an accurate uncertainty quantification (UQ) of a ZSINDy model.

Here we estimate the resolution parameter $\rho$ empirically from the residual between the observed derivative and the best fit ZSINDy model defined by the lowest free energy $f_\gamma$ and the coefficient values from \eqref{eqn:mu} (also known as Maximum A Posteriori). Since the resolution parameter is assumed to be the same across all dimensions of dynamics, we average it over the dimensions:
\begin{align}
	\rho_{est}=\sqrt{\frac{1}{n\cdot d}\sum\limits_{t,l}\left( \dot{x}_l-\vec{\Theta}^T_\gamma(\vec{x})\cdot \vec{\Xi}_{l,\gamma,MAP} \right)^2},
\end{align}
where the indices of the SINDy coefficients refer to the $l$th dimension, the selected set $\gamma$, and the MAP estimate (\eqref{eqn:mu}). The resulting $\rho_{est}$ values are reported in Fig.~6 of main text.

\subsection{Derivative error}
In order to understand the transition between the two behaviors of the residual, we consider the error of the numerical estimate of the derivative $\dot{x}$ where we leave out the dimension index for simplicity of notation. Fig.~5b of the main text shows that the inference breakdown is driven primarily by the error in the derivative (first term of the residual), and thus we focus on it. In this paper we use a simple central finite difference estimate, defined by the expression:
\begin{align}
	\dot{\tilde{x}}(t)\equiv\frac{x(t+\Delta t)-x(t-\Delta t)}{2\Delta t}=&
	\frac{x_0(t+\Delta t)-x_0(t-\Delta t)}{2\Delta t}+\frac{\delta x_{+} - \delta x_{-}}{2\Delta t}\nonumber\\
	=&\dot{x}_0(t)+\frac{\Delta t^2}{12}(\dddot{x\hspace{0pt}}_0(\tau_{-})+ \dddot{x\hspace{0pt}}_0(\tau_{+}))+\frac{\delta x_{+} - \delta x_{-}}{2\Delta t},\label{eqn:finitediff}
\end{align}
where $x_0$ refers to the true trajectory without noise, $\dot{x}_0$ to the true derivative, $\tau_{\pm}\in[t,t\pm\Delta t]$ are unknown evaluation points for the third derivative, and $\delta x_{\pm}$ are the realizations of the measurement noise at each of the two measurement points, known to be iid Gaussian with magnitude $\eta$.

In order to understand the variance of the derivative, we observe that while the evaluation points of the true third derivative are not known and generically vary with each time point $t$, they are uncorrelated from the noise realizations. The total variance of the derivative over a long time series then decomposes into two uncorrelated terms:
\begin{align}
	Var(\dot{x})=\left( \frac{\Delta t^2}{6} \right)^2 Var(\dddot{x\hspace{0pt}}_0)+\frac{2\cdot Var(\delta x)}{(2\Delta t)^2}=\order{\Delta t^4}+\frac{\eta^2}{2\Delta t^2},
\end{align}
which have different scaling. At small noise level, the derivative variance is dominated by the first term, the systematic error of the finite difference scheme. At large noise level, the noise driven second term dominates. This allows us to make a remarkably simple prediction of the resolution parameter in high noise limit:
\begin{align}
	\rho_{noise}=\eta/\sqrt{2}\Delta t,
	\label{eqn:rhoestnoise}
\end{align}
which we compare to the empirical estimation of the resolution parameter in Fig.~6c of the main text and find excellent agreement once the noise error overcomes the systematic error. Remarkably, while at noise above several percent the correct sparse set of coefficients is no longer identified, the same trend of the resolution parameter persists. In this extremely high noise regime the error in the residual is still dominated by the derivative rather than library, and is thus independent of the selected variable set so that \eqref{eqn:rhoestnoise} applies.

\subsection{Inference condensation}

How do we reconcile the persistent noise driven growth of the resolution parameter with the inference condensation shown in Fig.~3 of the main text?  The inference condensation shows that while the posterior mean of a particular coefficient quickly converges to a particular value $\mu$, the posterior standard deviation keeps decreasing with larger sample size. Specifically, for a generic coefficient we can write $\sigma=\sigma_0 \cdot(\rho/\rho_0)\sqrt{t_0/t}$, where $\rho_0,t_0$ are reference values for scaling.

For a library term to be explain the derivative at a statistically significant level, we typically require $\mu/\sigma>\order{1}$, i.e. the posterior mean is at least a few standard deviations away from zero. How much data is required to reach that condition? We can estimate the required trajectory length as follows:
\begin{align}
	\frac{\mu}{\sigma}>\order{1}\quad\Rightarrow\quad t>t_0 \rho^2 \frac{ \sigma_0^2}{\rho_0^2 \mu_0^2}\order{1},
\end{align}
which has quadratic scaling with the resolution parameter. In other words, if the noise in the data is increased by an order of magnitude, the required data for inference would grow by two orders of magnitude. The cost of such extensive data collection through experiments or simulations would quickly become prohibitive.

We thus showed that an empirical estimation of the resolution parameter $\rho$ allows for a valid uncertainty quantification across a wide range of noise levels, consistent with the observed inference condensation. Simultaneously, since the residual error is so strongly driven by the finite difference derivative error, denoising derivative approaches hold a large promise to improve SINDy tolerance to noise.

\subsection{Comparison with Ensemble SINDy}

Ensemble SINDy (E-SINDy), proposed recently by Fasel et al.~\cite{fasel2022ensemble}, is a powerful approach for dealing with noise and model uncertainty. The method estimates a distribution of the coefficient matrix $\Xi$ by sub-sampling the data and ensembling the resulting models. One of its most appealing advantages is its computational efficiency (in contrast to Bayesian SINDy~\cite{hirsh2022sparsifying}). 

\begin{figure*}[t]
	\centering
	\includegraphics[width=\textwidth]{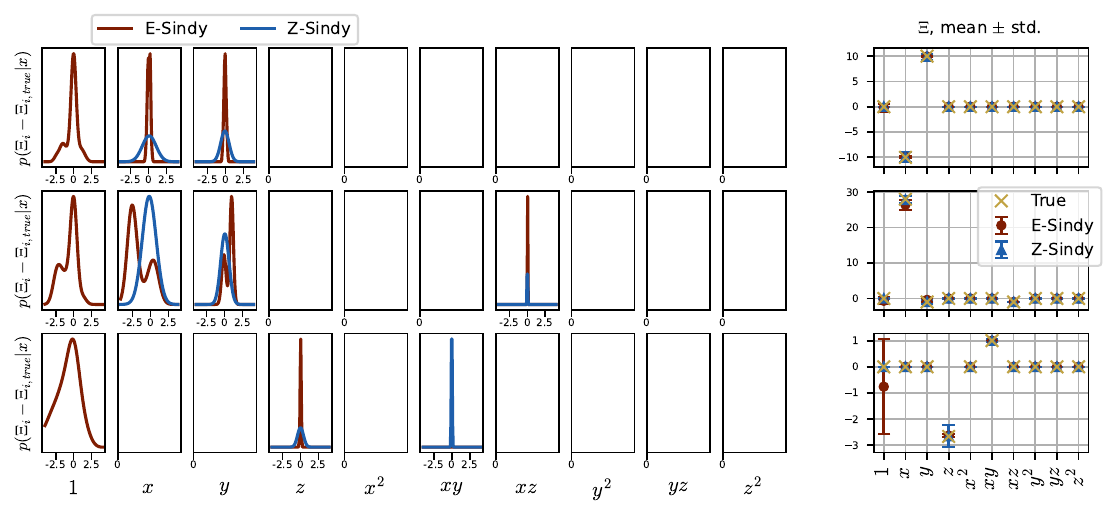}
	\caption{Comparison of posterior probability density functions of coefficients $p(\Xi_i - \Xi_{i,true} \vert x)$ between Ensemble SINDy and Z-SINDy, where $\Xi_i$ is the $i$-th row of $\Xi$ corresponding to the $i$-th variable. Missing distributions for either E-SINDy or Z-SINDy indicates that their corresponding optimizers always identify them correctly as zero.}
	\label{fig:evz}
\end{figure*}

In figure~\ref{fig:evz}, we compare the probability distribution of $\Xi$ in Z-SINDy with that of E-SINDy. For comparison, the hyperparameters were chosen to sparsity the models as much as possible in both methods. In this experiment, $\Delta t = 0.005$, $t_\text{end} = 20$, the Z-SINDy sparsity parameter per data point is $\lambda = 0.00125$, the threshold in E-SINDy is $0.8$, the noise magnitude is $\eta = 0.2$, the resolution parameter in Z-SINDy $\rho = \eta / (\Delta t\sqrt{2})$, and the number of ensembles in E-SINDy is $50$. The probability density functions of the E-SINDy coefficients were approximated using a Gaussian Kernel Density Estimator (KDE) with a Scott bandwidth estimator. 


The figure shows that Z-SINDy consistently identifies a sparser solution even when E-SINDy's sparsity coefficient (threshold) is increased to its maximum value (just below $1$, which is the smallest coefficient in the Lorenz System). This is expected since Z-SINDy more closely solves an $L_0$-norm problem compared to E-SINDy. Furthermore, some coefficient distributions are easier to interpret in Z-SINDy, such as the $x$ and $y$ terms in the $\dot y$ equation, given the potentially multi-modal distribution arising from E-SINDy. 

On the other hand, E-SINDy has narrower distributions in some cases, like the $x$ and $y$ terms in the $\dot x$ equation. This indicates that the $\rho=\eta / (\Delta t\sqrt{2})$ estimation of the resolution parameter that controls the width of the distribution might be an over-estimate for some of the coefficients. In this case, a combination of the two methods, such as estimating $\rho$ based on E-SINDy results can be explored in the future.





\section{Dynamical equations}
\chg{In this section we present the dynamical equations and parameter values for the Lorenz system used in the main text, as well as six additional systems.}

\subsection{Lorenz system}
\chg{The results in the main text are computed} for the classic Lorenz system \cite{lorenz1963deterministic} defined by the following set of ODEs:
\begin{align}
	\dot{x}=&\sigma(y-x)\\
	\dot{y}=&x(\rho-z)-y\\
	\dot{z}=&xy-\beta z,
\end{align}
where the ground truth parameter values are taken from the original paper \cite{lorenz1963deterministic} to be $\sigma=10$, $\beta=8/3$, $\rho=28$.

\subsection{Cubic oscillator}
\chg{The cubic oscillator system is defined by the following set of ODEs:}
\begin{align}
    \dot{x}&=ax^3+by^3\\
    \dot{y}&=cx^3+dy^3,
\end{align}
\chg{where the parameters are $a=-0.1,b=-2,c=2,d=-0.1$ as used in Ref.~\cite{fung2024rapid}. The solutions to this system are decaying nonlinear oscillations, which gradually increase in period. Notably, there is no linear part to the dynamics so linear analysis around the fixed point $(0,0)$ cannot capture any essentials of the dynamics.}

\subsection{Van der Pol oscillator}
\chg{The van der Pol oscillator is defined by the following set of ODEs:}
\begin{align}
    \dot{x}&=y\\
    \dot{y}&=by(1-x^2)-x,
\end{align}
\chg{where $b=4$. The solutions to this system approach a limit cycle.}

\subsection{Lotka-Volterra equations}
\chg{The Lotka-Volterra system is defined by the following set of ODEs:}
\begin{align}
    \dot{x}&=ax-bxy\\
    \dot{y}&=cxy-dy,
\end{align}
\chg{where $a=2/3,b=4/3,c=1,d=1$. The solutions to this system are a continuum of periodic trajectories selected by the initial conditions.}

\subsection{R\"ossler system}
\chg{The R\"ossler system is defined by the following system of ODEs:}
\begin{align}
    \dot{x}&=-y-z\\
    \dot{y}&=x+ay\\
    \dot{z}&=b+z(x-c),
\end{align}
\chg{where $a=0.2,b=0.2,c=5.7$. The solutions to this system is a chaotic trajectory with a single positive Lyapunov exponent.}


\subsection{Hyper R\"ossler attractor}
\chg{The Hyper R\"ossler system is a 4D extension of the regular R\"ossler system defined by the following system of ODEs:}
\begin{align}
    \dot{x}&=-y-z\\
    \dot{y}&=x+ay+w\\
    \dot{z}&=b+xz\\
    \dot{w}&=-cz+dw,
\end{align}
\chg{where the parameters $a=0.25,b=3,c=0.5,d=0.05$ result in a hyperchaotic trajectory with two positive Lyapunov exponents \cite{rossler1979equation, letellier2007hyperchaos}.}


\subsection{Abooee system}
\chg{The Abooee system is defined by the following system of ODEs:}
\begin{align}
    \dot{x}&=a(y-x)+byz^2\\
    \dot{y}&=cx+dxz^2\\
    \dot{z}&=hz+kx^2,
\end{align}
\chg{where the parameters $a=10,b=1,c=5,d=-1,h=-5,k=-6$ result in a chaotic trajectory \cite{abooee2013analysis}.}


\section{Additional systems}
\begin{figure}
	\centering
	\includegraphics[width=\textwidth]{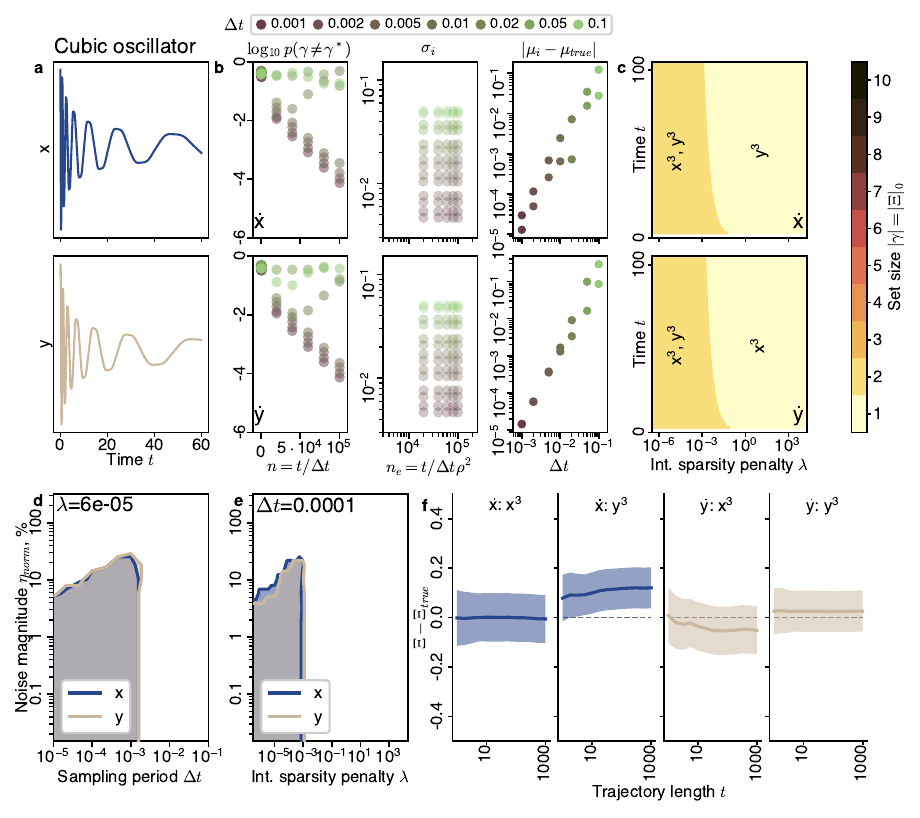}
	\caption{\chg{ZSINDy results for the cubic oscillator system. (a) Example of trajectories. (b) Inference condensation plots analogous to Fig.~3 of main text. (c) Sparsity promotion effect on the selected models analogous to Fig.~4b of main text. (d-e) Phase diagrams of the identifiable region, analogous to Fig.~7 of main text. (f) Confidence intervals of the coefficients on the correct model terms, analogous to Fig.~6d-e of the main text.}}
	\label{fig:cubic_oscillator}
\end{figure}

\begin{figure}
	\centering
	\includegraphics[width=\textwidth]{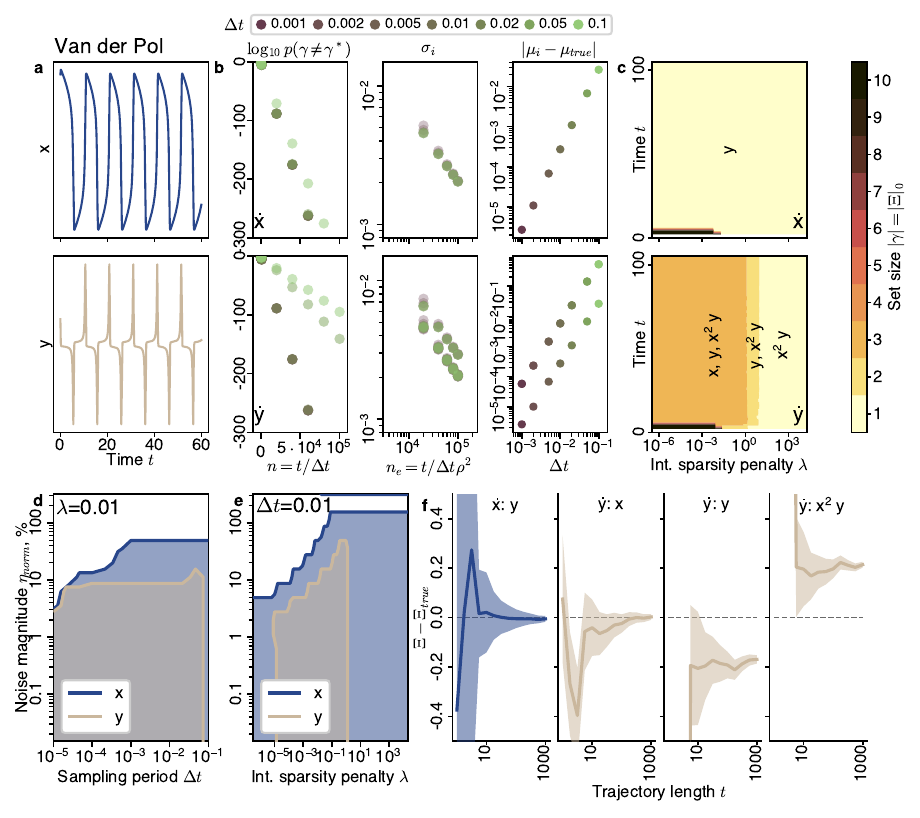}
	\caption{\chg{\chg{ZSINDy results for the van der Pol system. (a) Example of trajectories. (b) Inference condensation plots analogous to Fig.~3 of main text. (c) Sparsity promotion effect on the selected models analogous to Fig.~4b of main text. (d-e) Phase diagrams of the identifiable region, analogous to Fig.~7 of main text. (f) Confidence intervals of the coefficients on the correct model terms, analogous to Fig.~6d-e of the main text.}}}
	\label{fig:van_der_pol}
\end{figure}

\begin{figure}
	\centering
	\includegraphics[width=\textwidth]{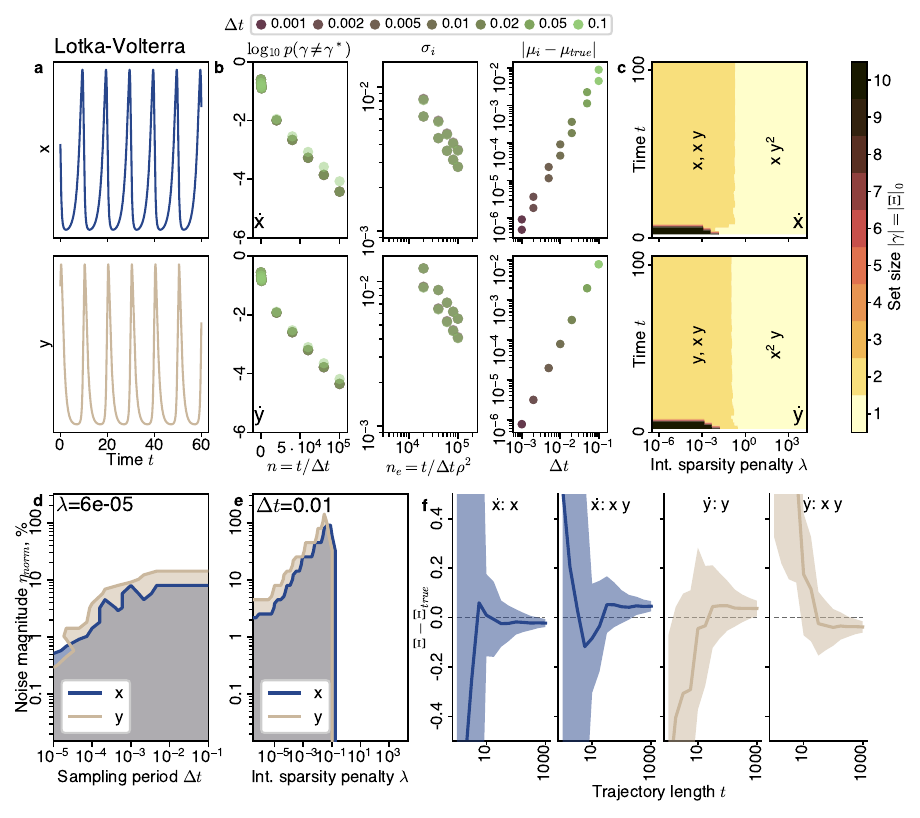}
	\caption{\chg{\chg{ZSINDy results for the Lotka-Volterra system. (a) Example of trajectories. (b) Inference condensation plots analogous to Fig.~3 of main text. (c) Sparsity promotion effect on the selected models analogous to Fig.~4b of main text. (d-e) Phase diagrams of the identifiable region, analogous to Fig.~7 of main text. (f) Confidence intervals of the coefficients on the correct model terms, analogous to Fig.~6d-e of the main text.}}}
	\label{fig:lotka_volterra}
\end{figure}

\begin{figure}
	\centering
	\includegraphics[width=\textwidth]{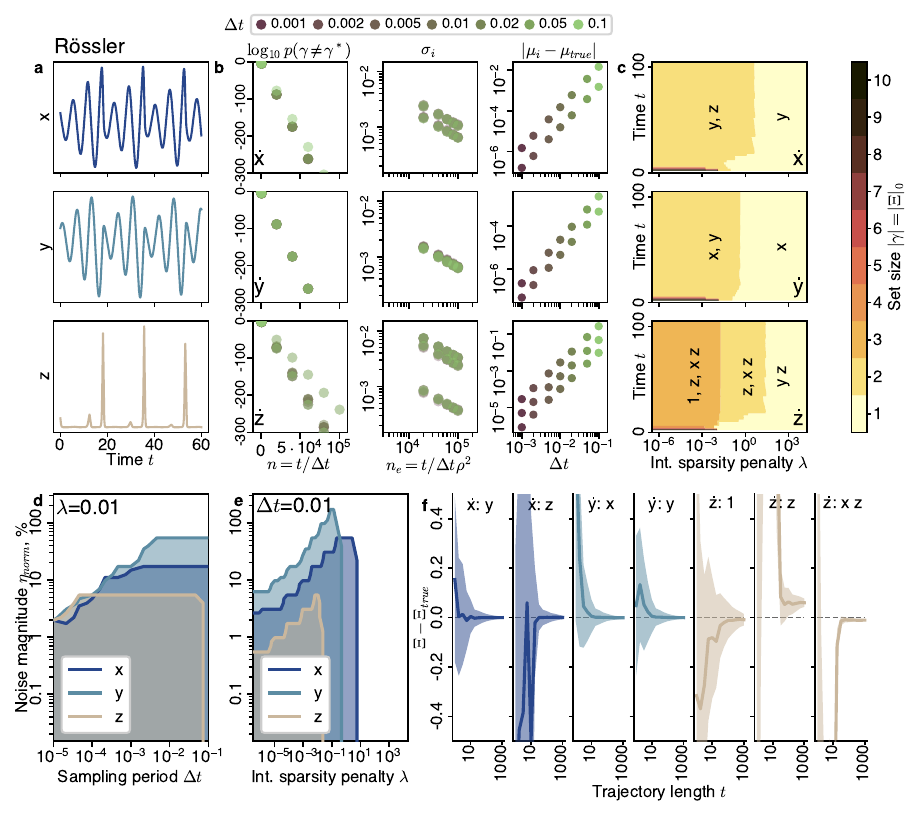}
	\caption{\chg{\chg{ZSINDy results for the R\"ssler system. (a) Example of trajectories. (b) Inference condensation plots analogous to Fig.~3 of main text. (c) Sparsity promotion effect on the selected models analogous to Fig.~4b of main text. (d-e) Phase diagrams of the identifiable region, analogous to Fig.~7 of main text. (f) Confidence intervals of the coefficients on the correct model terms, analogous to Fig.~6d-e of the main text.}}}
	\label{fig:rossler}
\end{figure}

\begin{figure}
	\centering
	\includegraphics[width=\textwidth]{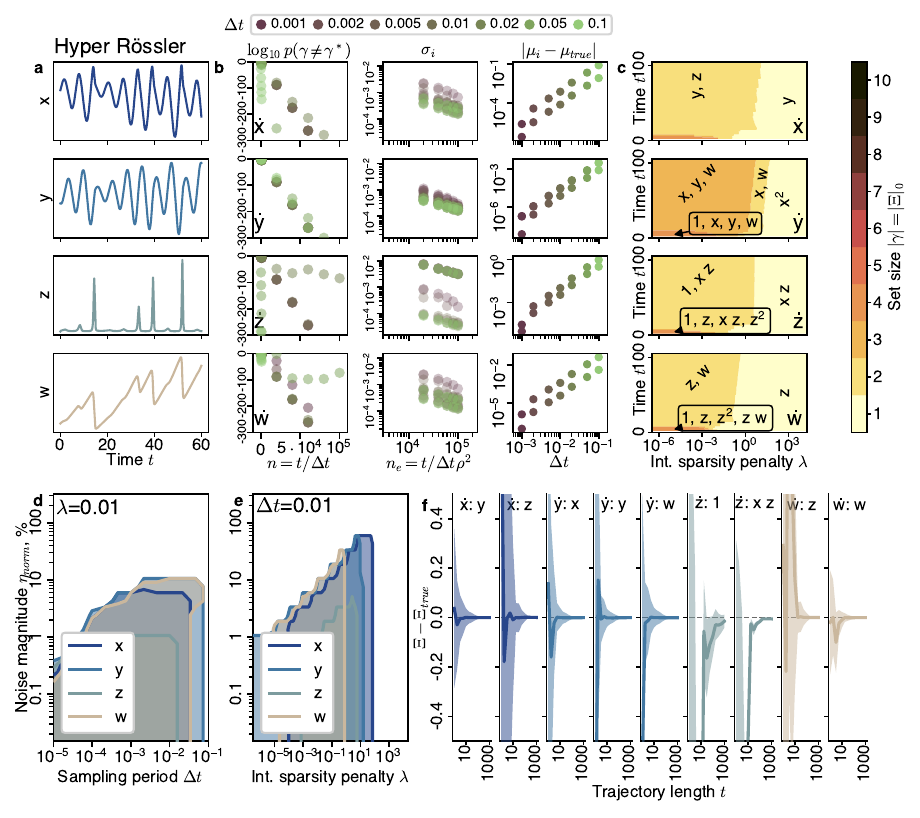}
	\caption{\chg{\chg{ZSINDy results for the Hyper R\"ossler system. (a) Example of trajectories. (b) Inference condensation plots analogous to Fig.~3 of main text. (c) Sparsity promotion effect on the selected models analogous to Fig.~4b of main text. (d-e) Phase diagrams of the identifiable region, analogous to Fig.~7 of main text. (f) Confidence intervals of the coefficients on the correct model terms, analogous to Fig.~6d-e of the main text.}}}
	\label{fig:hyper_rossler}
\end{figure}

\begin{figure}
	\centering
	\includegraphics[width=\textwidth]{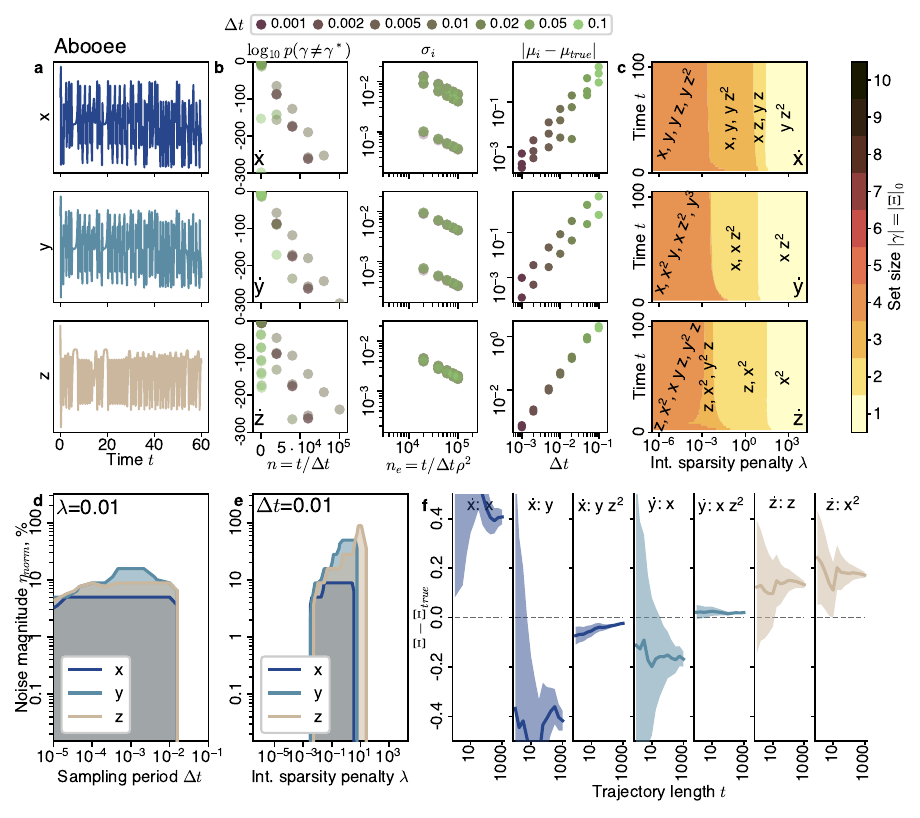}
	\caption{\chg{\chg{ZSINDy results for the Abooee system. (a) Example of trajectories. (b) Inference condensation plots analogous to Fig.~3 of main text. (c) Sparsity promotion effect on the selected models analogous to Fig.~4b of main text. (d-e) Phase diagrams of the identifiable region, analogous to Fig.~7 of main text. (f) Confidence intervals of the coefficients on the correct model terms, analogous to Fig.~6d-e of the main text.}}}
	\label{fig:abooee}
\end{figure}

\chg{For the additional systems, we follow largely the same analysis flow as for the Lorenz system in the main text. In particular, we report below the results for the cubic oscillator (Fig.~\ref{fig:cubic_oscillator}), van der Pol (Fig.~\ref{fig:van_der_pol}), Lotka-Volterra (Fig.~\ref{fig:lotka_volterra}), R\"ossler (Fig.~\ref{fig:rossler}), Hyper R\"ossler (Fig.~\ref{fig:hyper_rossler}), and Abooee (Fig.~\ref{fig:abooee}) systems. The qualitative trends of inference condensation, sparsity trade-off, phase diagrams, and uncertainty quantification remain mostly the same as for Lorenz, with a few exceptions listed below.}

\chg{We adjust the maximal polynomial power for the systems depending on their dimension and nature of terms. For the first four additional systems this still results in $N=10$ library terms and thus $2^N-1=1023$ sets considered ($\gamma=\varnothing$ is excluded), which can be enumerated exhaustively in a short time. For the last two additional systems the library size is larger, reaching $N=15$ for the 4D Hyper R\"ossler system and $N=20$ for the 3D Abooee system that has a higher order nonlinearity. For these two systems we perform set enumeration exhaustively up to a maximal number of variables included, 6 for Hyper R\"ossler and 5 for Abooee. This choice still enables us to identify the correct functional forms of the equations while confining computation to several seconds.}

\chg{Overall Z-SINDy is still able to identify the correct set of coefficients, even for more complex hyper-chaotic behavior or higher dimension or order of nonlinearity. For the van der Pol system, the true equation for $\dot{x}$ has a single term so it is correctly identified across the full range of sparsity penalties. Since we consider the same range of sampling period $\Delta t\in[10^{-5}, 10^{-1}]$ for uniform visualization of all systems, the right boundary of the correct identification phase (systematic error limited) is not always visible on (d) panels of respective figures.}

\chg{All but one systems show the same scaling of inference condensation: $p(\gamma\neq \gamma^*)\propto e^{-n}$, $\sigma_i\propto n^{-1/2}$, $\abs{\mu_i-\mu_{true}}\propto \Delta t^2$. Correspondingly, the error bars get smaller for longer trajectories, and eventually converge to a point that is systematically offset from the true answer. The key exception to this trend is the cubic oscillator system, for two reasons. First, as the oscillations in $x,y$ decay, the magnitude of the relevant terms $x^3,y^3$ decays much faster, so that later time points contribute much less to the covariances $C,\vec{V}$ and the covariances quickly saturate. As a result, the confidence interval in inference (panel f) remains finite width, while other systems show persistent dynamics, periodic or chaotic, and thus their confidence intervals shrink. Second, as the nonlinear cubic oscillations decay, unlike for linear oscillations, their frequency gets much lower. As a consequence, the competition between the sampling period $\Delta t$ and the characteristic time scale of the system is not constant, and the scaling of $\abs{\mu_i-\mu_{true}}$ is less clean.}

\chg{We simulated the additional systems using the same ranges of trajectory length, sampling periods, and noise magnitudes, but the underlying natural time scales of each system are different. This in particular results in different magnitudes of systematic errors. The right column on the panels b of each new system figure show the scaling of $\abs{\mu_i-\mu_{true}}$ in the \emph{noiseless} case with a clear $\order{\Delta t^2}$ trend. In contrast, the panels f show the statistical and systematic error under the constant \emph{absolute noise magnitude} $\eta=0.1$ for growing trajectory length. The asymptotic systematic error on noisy trajectory may either match the noiseless result (Lorenz, R\"ossler, Hyper R\"ossler, Abooee) or be quite a bit higher (cubic oscillator, van der Pol, Lotka-Volterra). The discrepancy might have to do with the chaotic nature of the former systems, or with the varying natural timescales of the systems, offering many further research questions.}








\section{Multi-trajectory fitting}
\begin{figure}
	\centering
	\includegraphics[width=\textwidth]{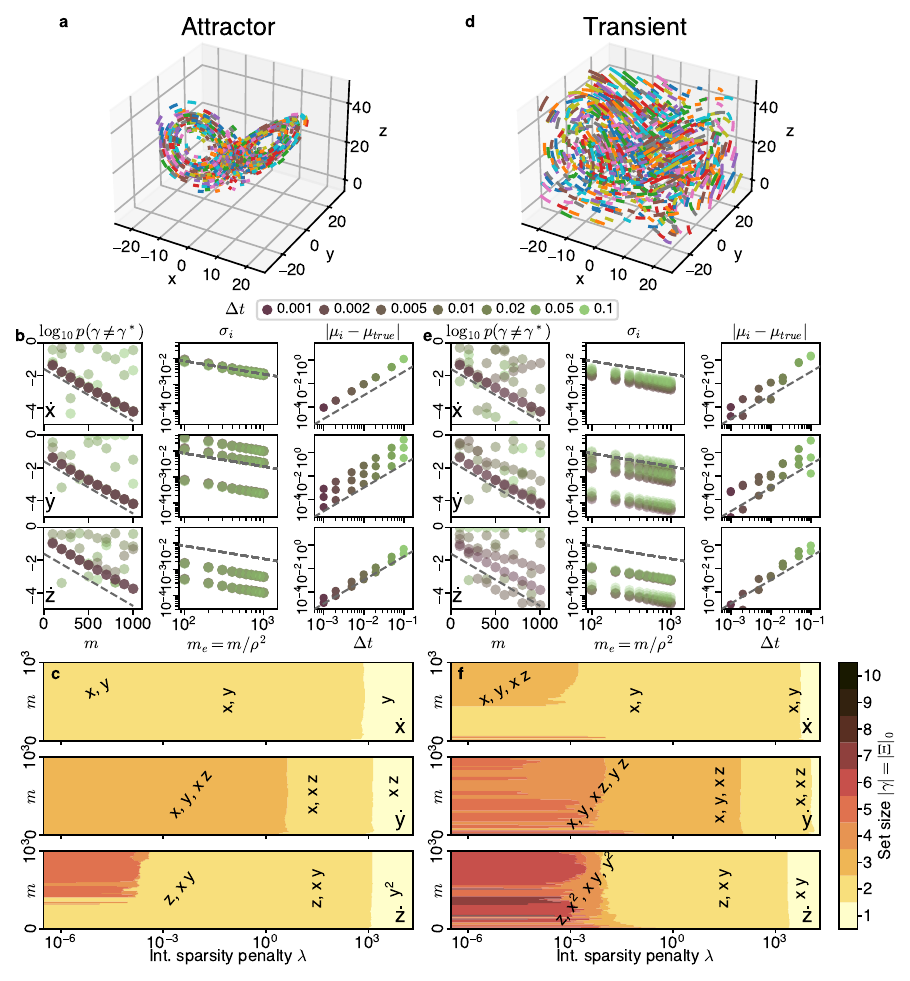}
	\caption{\chg{ZSINDy inference from multiple trajectories. Each trajectory has the length of 10 time steps with the initial conditions sampled either from the Lorenz attractor itself (a-c) or uniformly random points in a cuboid within 3$\sigma$ of the center of mass of the attractor (d-f). Each fit uses $m\in[1,1000]$ individual trajectories. (a,c) Examples of trajectories used for fitting. (b,e) Inference condensation plots analogous to Fig.~3 of main text. Within each plot type, the slope of the gray dashed lines and the axis bounds are consistent to facilitate comparison. (c,f) Sparsity promotion effect on the model form. Axes bounds and location of labels are consistent between the two columns to facilitate comparison.}}
	\label{fig:multitraj}
\end{figure}

\chg{We propose two setups for using multiple trajectories for ZSINDy fitting. In the first scenario, we integrate the Lorenz equations for a long time, discard the initial part of the trajectory, and then sample 1000 random points from the trajectory. In the second scenario, we compute the mean (center of mass) of the trajectory along with the standard deviations along the three coordinates. We then generate 1000 random points uniformly at random from the cuboid defined by $\textsf{mean}\pm 3\cdot\textsf{std}$ along each axis. Once the random points have been generated, we use them as initial conditions to integrate the Lorenz equations forward for 10 time steps with $\Delta t\in[0.001,0.1]$ (Fig.~\ref{fig:multitraj}a,d).}

\chg{Fig.~\ref{fig:multitraj}b,e presents the results of the multi-trajectory fit, where the number of trajectories $m\in[1,1000]$ takes the place of a single trajectory length. The condensation trends are generally similar with $p(\gamma\neq \gamma^*)\propto e^{-n}$, $\sigma_i\propto n^{-1/2}$, $\abs{\mu_i-\mu_{true}}\propto \Delta t^2$ regardless of the initial conditions sampling. The graphs for $p(\gamma\neq\gamma^*)$ have a lot more jitter for larger time steps $\Delta t$. White for larger time step the same number of 10 time steps covers a longer piece of trajectory, the numerical differentiation error grows, especially for the transient trajectories. On the contrary, the standard deviations $\sigma_i$ follow the same scaling but become smaller for a factor of 1.5-6 for the transient trajectories.}

\chg{The sparsity promotion plots in Fig.~\ref{fig:multitraj}c,f also look qualitatively the same as in the fits of continuous trajectories. Compared with the continuous trajectories, for attractor sampling the model boundaries shift to higher values of $\lambda$, and for transient sampling yet further higher. For transient sampling the inferred model is quite unstable for small values of $\lambda$ where sparsity promotion is weak but the few short trajectory bursts give inconsistent data.}




\section{Sampling frequency analysis}
\begin{figure}
	\centering
	\includegraphics[width=0.6\textwidth]{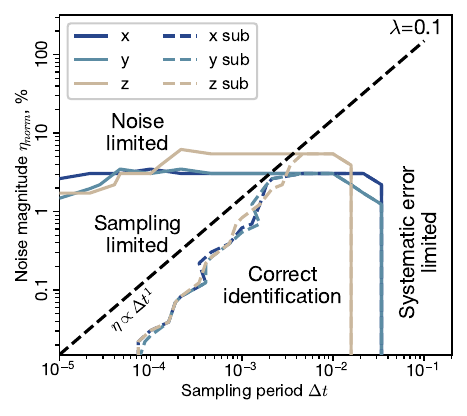}
	\caption{Inference of the correct dynamical equations is limited by the data noise $\eta$, sampling period $\Delta t$, and sampling density. The solid curves denote the region where the correct coefficient set is identified when every trajectory data point is used. The dashed curves denote the region where the data density per unit time $n/t$ is maintained as $\Delta t$ decreases. The region between the solid and dashed curves thus corresponds to the sampling limited regime, which disappears above $\Delta t \approx 5\cdot 10^{-3}$ where the curves for each dimension coincide. The black dashed curve shows the power law $\eta\propto \Delta t^1$ to guide the eye.}
	\label{fig:pd_subsample}
\end{figure}
The phase diagrams in Fig.~7 of main text showed that as the sampling period $\Delta t$ decreases, the boundary between the correct identification of the equation and the noise limited regime remains nearly flat. This shape results from an interplay of two opposite effects: while on one side the statistical error in the determination of derivative $\dot{\vec{x}}$ grows as $1/\Delta t$, on the other side the density of data points per unit time simultaneously grows as $1/\Delta t$, allowing averaging of the derivative and library terms over a large sample size. In order to disentangle these two effects, we subsample the trajectory to maintain a constant data density. We pick a reference density at $\Delta t_0=10^{-1}$ (10 data points per time unit). At each sampling period $\Delta t<\Delta t_0$, we retain only one in $\Delta t/\Delta t_0$ data points. The sampling period values $\Delta t$ decrease log-uniformly with a step of $10^{1/3}\approx 2.15$, thus for the progressively smaller periods the subsampling density is roughly 1 in 2 data points, then 1 in 5, 1 in 10, 1 in 20, 1 in 50 etc.

We compute the inference phase diagram for both the original and the subsampled data sets as shown in Fig.~\ref{fig:pd_subsample}. Without the benefit of increased data density, the region of correct identification shrinks with decreasing sampling period at a rate faster than $\eta\propto \Delta t$. At higher $\Delta t>5\cdot 10^{-3}$ the data set is still subsampled, but the identification boundary coincides with the one computed on the full data set. While further investigation would be required to determine the scaling at low $\Delta t$ more precisely, this example highlights the different meanings of ``data set size'' when applied to dynamical systems, specifically the issues of trajectory length vs sampling density.

\section{Computation time benchmark}

\chg{Here we present the results of benchmarking Z-SINDy computation. We ran the computation in single-thread form on the laptop Dell XPS 13 9340 with a CPU Intel Core Ultra 7 155H $\times$ 22. The reported time includes evaluating the free energies for all possible sets of coefficients for each dimension of dynamics, as well as evaluation of the mean and standard deviation for the best-fitting set, and is presented in Table~\ref{tab:benchmark}. The exact number of sets depends on both the dynamical system dimension and the maximal polynomial degree in the library. For five out of seven studied systems the enumeration was exhaustive, for the remaining two it was artificially limited to a maximal number of variables since we only wanted to consider sparse models. Since the code has deterministic runtime, the standard deviation across three replica runs is negligible. Most figures in the paper require running the Z-SINDy calculation multiple times to perform parameter sweeps. A key exception is the variation of the sparsity penalty $\lambda$, since after the non-sparsified intensive free energies $f'_\gamma$ are known, including sparsity penalty is computationally trivial addition $f_\gamma=f'_\gamma+\lambda \abs{\gamma}$.}

\chg{Following our own computational benchmark results, in Table~\ref{tab:time} we collate the qualitative method types and computational time reported in other Bayesian SINDy papers. We note that reporting computational hardware or runtime is \emph{de facto} not a standard practice in SINDy literature. In two cases, we got reports of computational time via personal communication. In particular, the algorithm of Hirsh et al.~\cite{hirsh2022sparsifying} took between 6-8 hours to fit the Mahaffy dataset of predator-prey interactions that has 21 data points (our Fig.~8 of main text). Where computational time is available, the iterative methods tend to be fast but only return the ``best'' models rather than the full landscape of models, whereas the Bayesian sampling methods take considerably longer and typically do not have convergence guarantees. We conclude that Z-SINDy is a computationally competitive method as compared to the literature.}

\begin{table}[]
    \centering
    \begin{tabular}{|l|c|c|c|c|c|c|c|}
    \hline
    System  &$d$	&max degree true   & max degree SINDy   &$N$    &max vars   &   number of sets  &compute time, sec  \\
    \hline
    Lorenz  &   3   &   2   &   2   &   10  &   10  &   1023$\times$3    &   0.305$\pm$0.007 \\
    Cubic oscillator    &   2   &   3   &   3   &   10  &   10  &   1023$\times$2    &   0.181$\pm$ 0.008    \\
    Van der Pol &   2   &   3   &   3   &   10  &   10  &   1023$\times$2    &   0.191$\pm$0.001 \\
    Lotka-Volterra  &   2   &   2   &   3   &   10  &   10  &   1023$\times$2    &   0.193$\pm$0.008 \\
    R\"ossler   &   3   &   2   &   2   &   10  &   10  &   1023$\times$3    &   0.280$\pm$0.002 \\
    Hyper R\"ossler &   4   &   2   &   2   &   15  &   6   &   9948$\times$4    &   3.52$\pm$0.04   \\
    Abooee  &   3   &   3   &   3   &   20  &   5   &   21699$\times$3   &   6.18$\pm$   0.04    \\
    \hline
    \end{tabular}
    \caption{\chg{Benchmarking runtime of Z-SINDy on the seven dynamical systems reported in the main text and SI. $d$ is the dimension of the dynamical system; max degree true is the highest polynomial degree that appears in the ground truth equations; max degree SINDy is the highest polynomial degree considered in the library; $N$ is the resulting number of library terms; max vars$\leq N$ is the largest number of items in the set that we consider; number of sets is the combinatorial number of sets considered for each dimension; compute time is reported in seconds with standard deviation across 3 replica runs.}}
    \label{tab:benchmark}
\end{table}

\begin{table}[]
    \begin{center}
    \begin{tabular}{|c|c|c|c|}
    \hline
    Ref.    &    Authors   &   Method  & Reported compute time \\
    \hline
        &   this manuscript & Z-SINDy  & 0.2-6 sec\\
    \cite{fung2024rapid}    &   Fung et al. 2024 & Greedy SparseBayes  & NA \\
    \cite{niven2024dynamical}    &   Niven et al. 2024 & JMAP/VBA  &  NA \\
    \cite{gao2022bayesian}    &   Gao et al. 2022 & autoencoder/SGLD  & NA/$\sim$30 sec$^*$  \\
    \cite{hirsh2022sparsifying}    &   Hirsh et al. 2022 & MCMC/NUTS  & NA/6-8 hours$^*$  \\
    \cite{north2022bayesian}    &   North et al. 2022 & MCMC  & NA  \\    
    \cite{1wang2020perspective}    & Wang et al. 2020  & MCMC/DRAM  &  NA \\
    \cite{3mathpati2024discovering}    & Mathpati et al. 2024  & VBA  & $\sim$2 sec  \\
    \cite{4tripura2023sparse}    & Tripura and Chakraborty 2023  & MCMC/Gibbs  &  NA \\
    \cite{5more2023bayesian}    & More et al. 2023 & VBA  &  NA \\
    \cite{6nayek2021spike}    & Nayek et al. 2021  & RVM/MCMC  & 0.03 sec / 3-30 sec  \\
    \cite{7long2024equation}    & Long et al. 2024  & EP-EM  & 100-4000 sec  \\
    \cite{8niven2020bayesian}    & Niven et al. 2020  & JMAP/VBA  &  NA \\
    \hline
    \end{tabular}
\end{center}
\caption{\chg{Reported computation times for Bayesian SINDy methods. $^*$ indicates the computation times reported through personal communication with the authors rather than article text. Computations are in most case performed on an average modern laptop or a dedicated workstation, as opposed to a cluster environment. NA=Not Available; JMAP=Joint Maximal A Posteriori; VBA=Variational Bayes Approximation; SGLD=Stochastic Gradient Langevin Dynamics; NUTS=No-U-Turn Sampler; MCMC=Markov Chain Monte Carlo; DRAM=Delayed Rejection Adaptive Metropolis; RVM=Relevance Vector Machine; EP-EM=Expectation-Propagation Expectation-Maximization}}
\label{tab:time}
\end{table}

\bibliography{biblio}